\documentclass[reprint,
amsmath,amssymb,
aps,
nofootinbib,
prc,
frontmatterverbose,
noeprint,
floatfix]{revtex4-1}
\usepackage{graphicx}
\usepackage{mathrsfs}
\usepackage{dcolumn}
\usepackage{bm}
\usepackage{cases}
\usepackage{float}
\usepackage{multirow,booktabs}
\usepackage{makecell}
\usepackage{amsthm}
\usepackage{epstopdf}
\usepackage{tabularx} 
\usepackage{graphicx}
\usepackage{dcolumn}
\usepackage{bm}
\usepackage{adjustbox}
\usepackage[mathlines]{lineno}

\usepackage{hyperref}
\hypersetup{
	colorlinks   = true, 
	linkcolor    = blue,  
	citecolor    = blue,  
	urlcolor     = blue,  
	filecolor    = blue,  
	pdfborder    = {0 0 0}
}



\begin{document}
	
	\title{Quantum-inspired Bayesian probability algorithm for nuclear mass predictions}
	
	\author{Kaizhong Tan$^{*}$\footnotemark[1]}
	\author{Jian Liu$^{*\,\dagger}$\footnotemark[1]\footnotemark[2]}
	\author{Chuan Wang}
	\affiliation{College of Science, China University of Petroleum (East China), Qingdao 266580, China}

	\begin{abstract}
		In this study, a novel quantum-inspired Bayesian probability (QIBP) algorithm, informed by quantum dynamics, is proposed to improve the predictions of nuclear mass from theoretical models. Within the QIBP framework, residuals between the theoretical and experimental mass values are mapped into wave functions in Hilbert space. The corresponding potentials are obtained by solving the Schr\"{o}dinger equation. Assuming that the residuals follow a Boltzmann distribution, the prior and likelihood probability density functions (PDFs) can be obtained from potentials. Finally, the Bayesian theorem is applied to derive the posterior PDF for estimating the target nuclear mass residuals. In global optimization, after employing the QIBP algorithm, the standard deviations of the WS4 model and the HFB model  with the SLy4 parameter set are reduced from 0.273 MeV and 5.250 MeV to 0.149 MeV and 0.324 MeV, respectively. In extrapolation analysis, the QIBP algorithm also effectively improves both models, indicating robust extrapolation capability. In addition, extrapolation based on the synthetic experimental set shows that the QIBP algorithm performs well near the known region and remains effective for most nuclides toward the drip lines. Furthermore, the QIBP algorithm is applied to predict $\alpha$-decay energies of Ra and Es isotopes, and the shell effects manifested in these isotopes are analyzed. This study validates the feasibility of quantum machine learning in nuclear mass research and demonstrates that the proposed algorithm can accurately describe nuclear masses, with potential applications in other areas of nuclear physics.

	\end{abstract}
	
	\maketitle
	
	\setcounter{footnote}{0}           
	\footnotetext[1]{The two authors contributed equally to this work.}
	\footnotetext[2]{Contact author: liujian@upc.edu.cn}

	\section{Introduction}
Machine learning (ML) has recently advanced at an unprecedented rate, achieving breakthroughs in various areas of fundamental science \cite{k1001, k1002, 1.e1, 1.e2}. In 2024, the Nobel Prize in Physics was awarded for the development of ML algorithms based on physical principles \cite{1.d1, k1005, k1006, RMP1, RMP2}, and the Nobel Prize in Chemistry honored the use of ML in the prediction of protein structure \cite{1.d2, k1007, k1008, k1009}. Recognizing the quantum nature of the world has prompted researchers to embed well-established classical theories into quantum Hilbert space. For example, Shannon’s information theory has evolved into quantum information theory \cite{k100b1}, and Turing’s theory of computation has given rise to quantum computing \cite{k100b2}. Like other classical theories, ML can also be reformulated within a quantum-mechanical framework. Embedding superposition \cite{1.e3}, entanglement \cite{1.e4}, and quantum dynamical features \cite{1.e5} into ML algorithms can overcome the computational bottlenecks of classical methods, enhance data mining capabilities, and discover novel solutions to specific problems. As a result, the field of quantum machine learning (QML) has emerged \cite{k10011}. It is widely regarded as a promising future direction for ML and is expected to provide powerful tools to accelerate scientific discovery \cite{k10012}. Particularly, QML is more likely to exhibit quantum advantage when handling data originating from quantum-mechanical processes, such as from experiments in physical chemistry \cite{k10015, k10016} and high-energy physics \cite{k10017, k10018, k10019, k10020, k10021}. Given that atomic nuclei are complex quantum many-body systems, combining QML with nuclear physics holds great potential.
	
There are mainly two categories of QML algorithms \cite{k10010}. The first class is designed to utilize quantum parallelism to accelerate the training of the ML model. It preserves the classical ML framework while implementing complex computations on quantum hardware. Representative examples include quantum principal component analysis \cite{k2004, k2005, k2006, k2007} and quantum support vector machines \cite{k2008, k2009, k20010}. The second class explores similarities between quantum systems and ML processes to develop new algorithms. The entire process of algorithms within this class can be implemented on classical computers, yet they offer more distinctive and advantageous solutions than existing ML algorithms. Such research is still limited, with notable works including the quantum clustering (QC) algorithm \cite{k20011, k20012} and the tensor networks \cite{k20013, kb002001}. Overall, different types of QML possess distinct advantages and can be flexibly applied based on specific problems.
	
The nuclear mass is the most fundamental property of nuclei, as it reflects the interactions among all constituents of the nucleus \cite{k3001}. It is crucial for nuclear and astrophysical studies to explore the limits of nuclear existence \cite{k3002, k3003}, reveal shell evolution \cite{k3004, 3.e1}, and understand the abundance of elements \cite{k3005, k3006}. Theoretical models for determining nuclear masses rely on different physical assumptions \cite{k3007, k3008, k3009, k30010, k30011, k30012}, such as spherical symmetry \cite{3.e2, k300b16} and local correlation \cite{k300b17, k30013}. In unknown regions far from the $\beta$-stability, the predictions of different models can differ by several to tens of MeV \cite{3.e3}. Hence, there is considerable scope for improving the existing nuclear mass models. In recent years, ML has emerged as a powerful tool for improving the nuclear mass predictions \cite{k1003, k1004}, with approaches including neural networks \cite{3.d1, 3.d2, k300b2, k300b3, k300b4, k300b5}, tree-based algorithms \cite{k300b8, k300b7}, Bayesian inference \cite{3.c1, 3.c2, k300b11}, and regression models \cite{k300b14, 9mgr-6mq7, RBF1, RBF2} achieving notable success. These data-driven methods identify complex relationships between nuclear mass data and relevant features, and capture subtle patterns and correlations that might be overlooked by theoretical models, thus exhibiting superior generalization capabilities. Many studies have reduced the standard deviation in nuclear mass predictions to below 200 keV, and several newly proposed magic numbers have been validated \cite{k300b6}. These advancements offer crucial support for studies of exotic nuclear structures and the \textit{r}-process \cite{k300b15}.
	
Both the theoretical description and the experimental determination of nuclear masses are closely related to quantum mechanics \cite{4.e1, 4.e2}. Compared with classical ML, QML can describe quantum systems more naturally and is therefore suited to nuclear mass predictions. This study proposes a quantum-inspired Bayesian probability (QIBP) algorithm to improve nuclear mass predictions. In the QIBP framework, the residuals $\delta$ between theoretical and experimental mass values are treated as states in Hilbert space and mapped to wave functions $\psi$. By treating $\psi$ as eigenstates of the Schr\"{o}dinger equation, the potentials are derived through simple analytical operations. Assuming that the residuals follow a Boltzmann distribution, the prior and likelihood probability density functions (PDFs) can be obtained from the Schr\"{o}dinger potentials. Finally, the posterior PDF is determined via Bayesian theory, and its expectation value is evaluated to provide a precise estimate of the residual $\delta$, which is then used to optimize the theoretical models.
	
In this study, the QIBP algorithm is applied to two representative theoretical models, the macroscopic-microscopic WS4 model \cite{k6002} and the microscopic Hartree-Fock-Bogoliubov (HFB) model with the SLy4 parameter set \cite{k6001}. Global optimization and extrapolation analyses are performed to evaluate the capability of the QIBP algorithm to optimize theoretical models and to capture quantum effects and other subtle patterns. In addition, the advantages of the QML algorithm are discussed through a comparison with classical Bayesian inference methods. Finally, $\alpha$-decay energies of Ra and Es isotopes are predicted using the QIBP algorithm, and subshell effects exhibited by the Es isotopes are analyzed. Results show that the QIBP algorithm can provide accurate predictions for nuclear masses of unknown nuclei, with potential applications in other areas of nuclear physics, such as nuclear reactions and astrophysics.
	
The structure of this paper is as follows: Sec. \ref{sec:2} discusses the theoretical framework of the QIBP algorithm; Sec. \ref{sec:3} presents the results and related analyses of the QIBP algorithm; and Sec. \ref{sec:4} provides the conclusion.
	
\section{Theoretical framework}\label{sec:2}
	
This section introduces the theoretical framework of the quantum-inspired Bayesian probability (QIBP) algorithm. The QIBP algorithm is designed to predict the residuals between theoretical and experimental nuclear mass values, thereby improving the performance of theoretical models. Its structure consists of two principal components: an external framework constructed using Bayesian theory and a core module for probability density estimation informed by quantum dynamics.
	
\subsection{The multivariate Bayesian theorem}\label{sec:2A}
	
The external framework of the QIBP algorithm is based on the multivariate Bayesian theorem, which forms its classical part. In the QIBP algorithm, the residual $\delta$ of nuclear mass is treated as a continuous variable. For a target nucleus with proton number $Z_t$ and neutron number $N_t$, it is assumed that $Z_t$ and $N_t$ are independent features. According to the multivariate Bayesian theorem, the posterior probability density function (PDF) of $\delta$ for the target nucleus $(Z_t,N_t)$ can be expressed as
	
	\begin{equation}
		p(\delta|Z_t,N_t)=\frac{p(Z_t|\delta)\,p(N_t|\delta)\,p(\delta)}{\displaystyle \int p(Z_t|\delta)\,p(N_t|\delta)\,p(\delta)\,\mathrm{d}\delta}.
		\label{E1}
	\end{equation}
	
	The likelihood PDFs $p(Z_t|\delta)$ and $p(N_t|\delta)$ in Eq.~(\ref{E1}) can be obtained by the univariate Bayesian formula,
	\begin{equation}
		p(Z_t|\delta)=\frac{p(\delta|Z_t)p(Z_t)}{p(\delta)},
		\label{E2}
	\end{equation}
	\begin{equation}
		p(N_t|\delta)=\frac{p(\delta|N_t)p(N_t)}{p(\delta)}.
		\label{E3}
	\end{equation}
	The prior probabilities $p(Z_t)$ and $p(N_t)$ depend on the frequencies of occurrence of nuclei with the same proton number $Z_t$ and the neutron number $N_t$ in the dataset, respectively. Note that if the dataset lacks nuclides with the same proton or neutron number as the target nucleus, the likelihood PDF $p(Z_t|\delta)$ or $p(N_t|\delta)$ is treated as a constant in Eq.~(\ref{E1}). Since the posterior PDF is normalized afterward, the choice of this constant does not affect the final result. To calculate the posterior PDF of $\delta$, it is sufficient to determine three PDFs, the prior PDF $p(\delta)$, and two likelihood PDFs, $p(\delta|Z_t)$ and $p(\delta|N_t)$.

	\subsection{Quantum-inspired determination of probability density functions}
	
	The core of the QIBP algorithm is applying concepts from quantum dynamics, including wave functions and the Schr\"{o}dinger equation, to determine the three PDFs described in Sec. \ref{sec:2A}. The residuals are first mapped to wave functions in Hilbert space, from which the corresponding potentials are obtained by solving the Schr\"{o}dinger equation. Assuming that the residuals follow a Boltzmann distribution, the PDFs can be derived from the potentials. Specifically, for a dataset containing $n$ residuals, the wave function $\Psi(\delta)$ for the target nucleus $(Z_t, N_t)$ is defined using Gaussian kernels as
	
	\begin{equation}
		\Psi(\delta)=\sum_{i=1}^{n} e^{-\frac{(\delta-\delta_i)^2}{2\lambda^{2}}},
		\label{E4}
	\end{equation}
	where $\delta_i=\delta^{\rm exp}_i-\delta^{\rm th}_i$ represents the raw residual of each nucleus in the dataset. $\lambda$ is the bandwidth parameter, whose value is set to the standard deviation of $\delta_i$, and it characterizes the spread of the data distribution. To effectively predict nuclear masses, a weight coefficient $w_{i,t}$ is introduced to capture local features,
	
	\begin{equation}
		w_{i,t}=\exp\!\left[-\frac{(Z_i-Z_t)^2+(N_i-N_t)^2}{2}\right]+\epsilon.
		\label{E5}
	\end{equation}
	The parameter $\epsilon$ affects the stability of the algorithm, and we set $\epsilon=10^{-10}$. After introducing the weight coefficients, the wave function is expressed as
	\begin{equation}
		\psi_t(\delta)=\sum_{i=1}^{n}w_{i,t}e^{-\frac{(\delta-\delta_i)^2}{2\lambda^{2}}}.
		\label{E6}
	\end{equation}
	
	To determine the likelihood PDFs $p(\delta|Z_t)$ and $p(\delta|N_t)$, wave functions associated with the isotopic and isotonic chains are also derived using the same procedure,
	
	\begin{equation}
		\psi_{Z_t}(\delta)=\sum_{i=1}^{n_Z}w_{i,t}e^{-\frac{(\delta-\delta_i)^2}{2\lambda^{2}}},
		\label{E7}
	\end{equation}
	\begin{equation}
		\psi_{N_t}(\delta)=\sum_{i=1}^{n_N}w_{i,t}e^{-\frac{(\delta-\delta_i)^2}{2\lambda^{2}}}.
		\label{E8}
	\end{equation}
	Here, $n_Z$($n_N$) represents the number of nuclei with $Z_t$($N_t$). When constructing $\psi_{Z_t}(\delta)$ and $\psi_{N_t}(\delta)$, only the nuclei in the dataset that have the same proton number $Z_t$ or the same neutron number $N_t$ as the target nucleus are considered. The QIBP algorithm treats $\psi_t(\delta)$, $\psi_{Z_t}(\delta)$, and $\psi_{N_t}(\delta)$ as eigenstates of the Schr\"{o}dinger equation,
	\begin{equation}
		H\psi=\left[-\frac{\lambda^{2}}{2}\,\frac{\mathrm{d}^{2}}{\mathrm{d}\delta^{2}}+V\right]\psi=E\psi.
		\label{E9}
	\end{equation}
	Here we rescaled $H$ and $V$ of the conventional quantum-mechanical equation to leave only one parameter $\lambda$. The potentials $V_t(\delta)$, $V_{Z_t}(\delta)$, and $V_{N_t}(\delta)$ are derived by solving Eq.~\eqref{E9},
	\begin{equation}
		V_t(\delta)=E_t-\frac{1}{2}+\frac{1}{2\lambda^{2}\psi_t(\delta)}\sum_{i=1}^{n}w_{i,t}(\delta-\delta_i)^{2}e^{-\frac{(\delta-\delta_i)^2}{2\lambda^{2}}},
		\label{E10}
	\end{equation}
	\begin{equation}
		V_{Z_t}(\delta)=E_{Z_t}-\frac{1}{2}+\frac{1}{2\lambda^{2}\psi_{Z_t}(\delta)}\sum_{i=1}^{n_Z}w_{i,t}(\delta-\delta_i)^{2}e^{-\frac{(\delta-\delta_i)^2}{2\lambda^{2}}},
		\label{E11}
	\end{equation}
	\begin{equation}
		V_{N_t}(\delta)=E_{N_t}-\frac{1}{2}+\frac{1}{2\lambda^{2}\psi_{N_t}(\delta)}\sum_{i=1}^{n_N}w_{i,t}(\delta-\delta_i)^{2}e^{-\frac{(\delta-\delta_i)^2}{2\lambda^{2}}}.
		\label{E12}
	\end{equation}
	By further requiring that the minimum value of $V$ is zero, we obtain
	\begin{equation}
		E=-\min\frac{\lambda^{2}}{2\psi}\frac{\mathrm{d}^{2}\psi}{\mathrm{d}\delta^{2}},
		\label{E13}
	\end{equation}
	thereby uniquely determining $E_t$, $E_{Z_t}$, and $E_{N_t}$ in Eqs.~\eqref{E10}-\eqref{E12}.
	
	As a prominent QML method, the quantum clustering (QC) algorithm was the first to employ such potentials for data classification and conducted an in-depth analysis of their robustness. From a physical perspective, the minima of potential act as abstract gravitational sources, attracting the distribution of data toward more stable (lower-energy) regions. Therefore, in the QC algorithm, the minima of the potential are taken as cluster centers, and a gradient-descent method is used to assign each data point to a cluster. Since the potential contains information from both the second derivative and the gradient of the wave function $\psi$, it offers greater sensitivity and robustness in data processing. The potential thus captures the subtle features of the data distribution.
	
	The purpose of this work is not to perform clustering through the potential. Instead, the potential is used to extract the probability density distribution of $\delta$. The well-known Boltzmann machine employs the mathematical form of the Boltzmann distribution to derive probability density functions from an energy function \cite{ACKLEY1985147, EBM}. This study adopts a similar approach. Assuming that the residuals follow a Boltzmann distribution, the prior PDF~$p(\delta)$ and the likelihood PDFs~$p(\delta|Z_t)$ and~$p(\delta|N_t)$ can be derived from the corresponding potentials,
	\begin{align}
		p(\delta)      &= \frac{1}{C_t}\,e^{-\beta V_t(\delta)},                              \label{E14}\\[4pt]
		p(\delta|Z_t)  &= \frac{1}{C_{Z_t}}\,e^{-\beta_{Z}V_{Z_t}(\delta)},                      \label{E15}\\[4pt]
		p(\delta|N_t)  &= \frac{1}{C_{N_t}}\,e^{-\beta_{N}V_{N_t}(\delta)},                      \label{E16}
	\end{align}
	where $C_t$, $C_{Z_t}$, and $C_{N_t}$ normalize the three PDFs, and the smoothing parameters $\beta$, $\beta_Z$, and $\beta_N$ affect the stability of the PDFs, respectively. By substituting the three PDFs into Eqs.~\eqref{E1}-\eqref{E3} of the Bayesian framework, the posterior PDF $p(\delta|Z_t,N_t)$ for the target nucleus $(Z_t,N_t)$ is obtained.  The final estimated residual for the target nucleus is then determined by calculating the expectation value,
	\begin{equation}
		\delta^{\mathrm{em}}(Z_t,N_t)=\int \delta\,p(\delta|Z_t,N_t)\,\mathrm{d}\delta.          \label{E17}
	\end{equation}
	
	Finally, the refined mass prediction $E^{\mathrm{corr}}(Z_t,N_t)$ is obtained by adding the estimated residual to the theoretical mass $E^{\mathrm{th}}(Z_t,N_t)$,
	
	\begin{equation}
		E^{\mathrm{corr}}(Z_t,N_t)=E^{\mathrm{th}}(Z_t,N_t)+\delta^{\mathrm{em}}(Z_t,N_t). \label{E18}
	\end{equation}
	
	The standard deviation $\sigma_{\mathrm{rms}}$ is used to quantify the deviation between the refined predictions and the experimental data. It is defined as
	\begin{equation}
		\sigma_{\mathrm{rms}}^{2}=\frac{1}{n}\sum_{i=1}^{n}\bigl(E^{\mathrm{corr}}_{i}-E^{\exp}_{i}\bigr)^{2}. \label{E19}
	\end{equation}
	
	In the QIBP algorithm, the predicted uncertainty is derived from the posterior PDF.  The one-sigma uncertainty associated with the prediction for the nucleus $(Z_t,N_t)$ is given by
	\begin{equation}
		\Delta E(Z_t,N_t)=\sqrt{\int\!\bigl[\delta-\delta^{\mathrm{em}}(Z_t,N_t)\bigr]^{2}
			p(\delta|Z_t,N_t)\,\mathrm{d}\delta}. \label{E20}
	\end{equation}

	\section{Results and discussion}\label{sec:3}
	
This section presents the performance of the QIBP algorithm in improving nuclear mass predictions. AME2020 compiles the nuclear masses of 3557 nuclides, of which 2457 are determined experimentally and 1100 are estimated \cite{k1}. To ensure the rigor of this study, only nuclides with experimentally measured masses are considered. A total of 2253 nuclei with proton number $Z \ge 20$ and neutron number $N \ge 20$ from AME2020 are selected as the entire set. Theoretical models employed in this study include the macroscopic-microscopic WS4 model \cite{k6002} and the microscopic Hartree-Fock-Bogoliubov (HFB) model with the SLy4 parameter set \cite{k6001}. The theoretical mass values of 2253 nuclei are first calculated from these models, and the corresponding residuals are determined as $\delta^{\mathrm{pre}} = E^{\exp} - E^{\mathrm{th}}$. The QIBP algorithm is then applied to optimize the predictions of each model. Global optimization and extrapolation analyses are performed to assess the effectiveness of the QIBP algorithm in refining theoretical models and capturing subtle patterns. Finally, the predictive capability of the QIBP algorithm for $\alpha$-decay energies is demonstrated.
	
In determining the three PDFs $p(\delta)$, $p(\delta|Z_t)$, and $p(\delta|N_t)$, there are three undefined parameters, $\beta$, $\beta_{Z}$, and $\beta_{N}$. In general, $p(\delta)$ describes the global distribution of nuclear mass residuals and should be modeled smoothly to mitigate noise. $p(\delta|Z_t)$ and $p(\delta|N_t)$ describe local residual distributions and typically require finer resolution. To ensure the stability of the QIBP algorithm, a basic guideline is that $\beta$ should be smaller than $\beta_{Z}$ and $\beta_{N}$. Under this condition, the QIBP algorithm achieves robust optimization performance. When $\beta_{Z} = \beta_{N} = 1.0$, variations of $\beta$ from 1/2.0 to 1/1.1 exert negligible influence on the prediction results. Therefore, the general parameter combination for the QIBP algorithm is set as $\beta = 1/1.2$, $\beta_{Z}=1.0$, and $\beta_{N}=1.0$.
	
\subsection{Global optimizations of the QIBP algorithm}\label{III A}

\begin{figure*}[htpb]
		\centering
		\includegraphics[width=18cm,angle=0,clip=true]{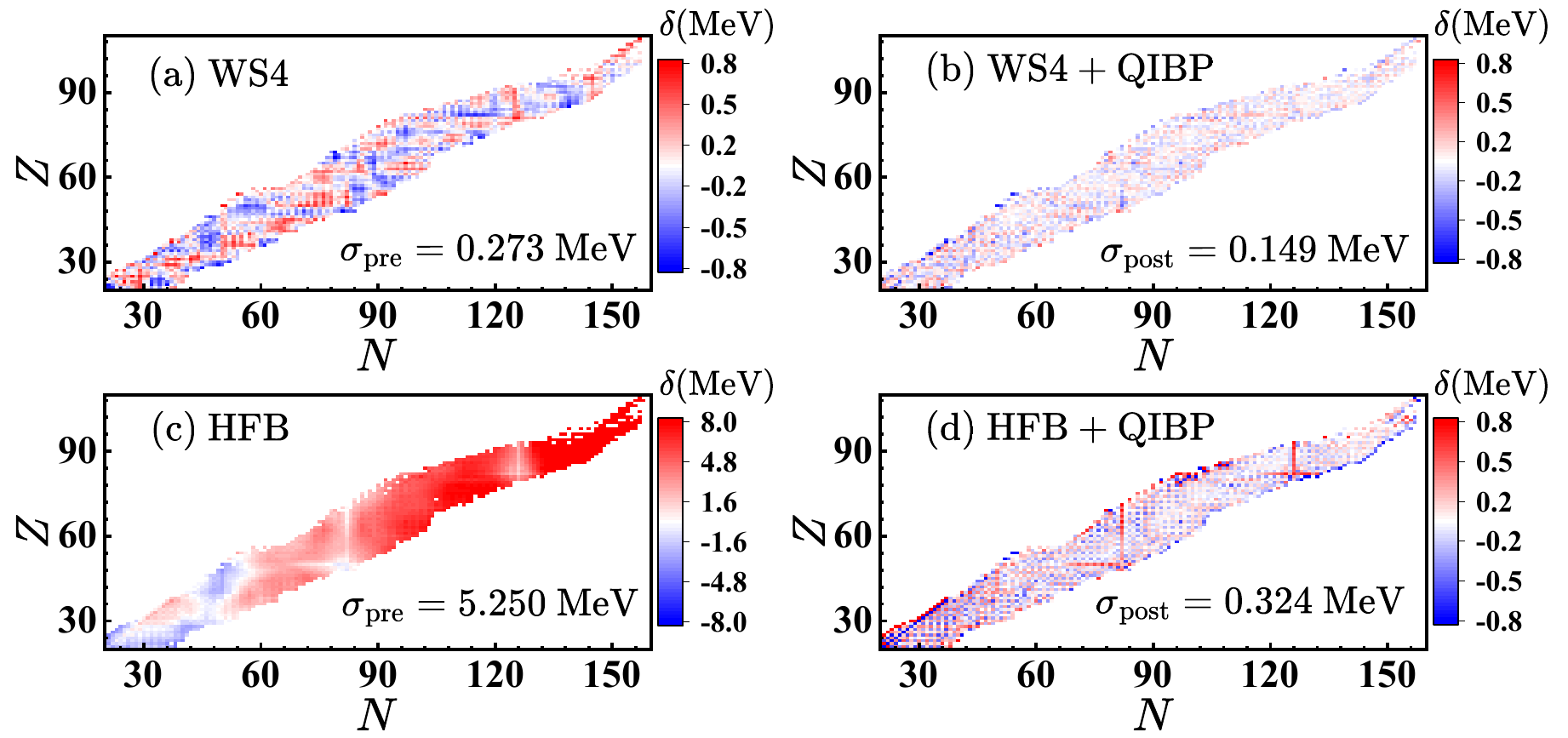}\\
		\caption{(a) The raw nuclear mass residuals $\delta$ (MeV) from the WS4 model. (b) The nuclear mass residuals from the WS4 model after refinement by the QIBP algorithm. (c) The same as (a), but for the HFB model. (d) The same as (b), but for the HFB model after refinement by the QIBP algorithm.}
		\label{fig1}
	\end{figure*}

Leave-one-out cross-validation is employed in global optimization analysis. Specifically, when predicting the mass of a target nucleus, the remaining 2252 nuclei in the entire set are used as the training set. Following the procedure described in Sec. \ref{sec:2}, the wave functions $\psi$ are determined by Eqs.~\eqref{E6}-\eqref{E8}, and the corresponding potentials $V$ are obtained from Eqs.~\eqref{E10}-\eqref{E12}. Within the Bayesian framework, the posterior PDF $p(\delta|Z_t, N_t)$ is determined by integrating Eqs.~\eqref{E1}-\eqref{E3}. The refined nuclear mass is finally calculated using Eqs.~\eqref{E17} and~\eqref{E18}. The same calculation is applied to each nucleus in the entire set. To demonstrate the optimization performance of the QIBP algorithm, Table~\ref{T1} presents the standard deviations before and after optimization, denoted as $\sigma^{\mathrm{pre}}$ and $\sigma^{\mathrm{post}}$, respectively. The improvement rates achieved by the QIBP algorithm are quantified as $\Delta\sigma/\sigma^{\mathrm{pre}}=(\sigma^{\mathrm{pre}}-\sigma^{\mathrm{post}})/\sigma^{\mathrm{pre}}$. In addition, the distributions of the residuals before and after optimization across the nuclear chart are illustrated in Fig.~\ref{fig1}.

\begin{table}[ht]
	\centering
	\caption{Standard deviations $\sigma_{\text{pre}}$ (MeV) from the WS4 and HFB models, as well as $\sigma_{\text{post}}$ (MeV) after QIBP and CBP refinements. A total of 2253 nuclei in AME2020 with $Z\ge20$ and $N\ge20$ are chosen as the entire set.}
	\label{T1}
	
	\renewcommand\arraystretch{1.2}
	\setlength{\tabcolsep}{8pt}
	
	\begin{tabular*}{\linewidth}{@{\extracolsep{\fill}} l @{\hspace{0.5em}} c c @{\hspace{0.4em}} c @{\hspace{0em}} >{\hspace{0.6em}} c @{\hspace{0.4em}} c @{\hspace{0.1em}}}
		\hline\hline
		\multirow{2}{*}{Models} & \multirow{2}{*}{$\sigma_{\text{pre}}$}
		& \multicolumn{2}{c}{QIBP} & \multicolumn{2}{c}{CBP} \\
		\cline{3-4}\cline{5-6}
		&  & $\sigma_{\text{post}}$ & $\Delta\sigma/\sigma_{\text{pre}}$
		& $\sigma_{\text{post}}$ & $\Delta\sigma/\sigma_{\text{pre}}$ \\
		\hline
		WS4 & 0.273 & 0.149 & 45.4\% & 0.187 & 31.5\% \\
		HFB & 5.250 & 0.324 & 93.8\% & 0.509 & 90.3\% \\
		\hline\hline
	\end{tabular*}
\end{table}

Table~\ref{T1} demonstrates that the QIBP algorithm enhances the performance of both theoretical models. For the WS4 model, the standard deviation is reduced from $\sigma_{\text{pre}}$ = 0.273 MeV to $\sigma_{\text{post}}$ = 0.149 MeV, corresponding to an improvement rate of 45.4\%. In addition to the improvement in overall standard deviation, changes in the residual distribution across the nuclear chart can further demonstrate the optimization capability of the QIBP algorithm. As shown in Fig.~\ref{fig1}(a), before applying the QIBP algorithm, patterns in the residuals distribution remain observable. Some regions show clusters of positive residuals, indicated by concentrated red areas, while others are dominated by negative residuals, appearing as blue zones. This clustering pattern of residuals indicates that certain physical effects are still not incorporated into the WS4 model. In addition, the residuals exhibit a certain level of odd-even staggering, suggesting that this effect is not fully captured by the WS4 model. After applying the QIBP algorithm, the residuals are reduced across the entire nuclear chart. The distribution of residuals becomes more random, and the odd-even staggering is weakened. This suggests that the QIBP algorithm effectively incorporates patterns and correlations overlooked by the WS4 model, resulting in a lower standard deviation.
	
For the HFB model, the standard deviation is reduced from $\sigma^{\text{pre}} = 5.250\ \text{MeV}$ to $\sigma^{\text{post}} = 0.324\ \text{MeV}$, corresponding to an improvement rate of 93.8\%. In Fig.~\ref{fig1}(c), the residuals of the HFB model before optimization are large and exhibit a clearer pattern. In regions near proton and neutron magic numbers, the mass deviations are significantly smaller than those in surrounding areas. This is because the HFB model with the SLy4 parameter set is solved under the assumption of spherical symmetry and does not account for deformation effects. As a result, the HFB model provides accurate descriptions of doubly magical nuclei, but shows systematic deviations for deformed nuclei.  Fig.~\ref{fig1}(d) illustrates that the QIBP algorithm identifies correlations in mass residuals and effectively captures deformation effects not included in the HFB model, thereby enabling improved descriptions of nuclear mass. It should be noted that the color scales used in Fig.~\ref{fig1}(c) and Fig.~\ref{fig1}(d) differ by one order of magnitude. Although some relatively dark regions still appear in Fig.~\ref{fig1}(d) after refinement with the QIBP algorithm, for example in the vicinity of magic numbers, the results remain significantly improved compared with the original predictions of the HFB model.
	
The above discussion indicates that, building on the global descriptions provided by theoretical models, the QIBP algorithm can further capture patterns that are not taken into account by these models, thereby achieving effective refinement. The performance of the QIBP algorithm originates from its theoretical framework. As a QML approach, the PDFs are calculated using the Schr\"{o}dinger equation in quantum mechanics, which serves as a more advanced mathematical tool. These PDFs exhibit strong robustness and are capable of accurately capturing underlying subtle effects and correlations from nuclear mass residual data, such as odd-even staggering and deformation. In addition, the Bayesian framework, together with the weight coefficients, incorporates statistical correlations among neighboring nuclei, all of which contribute to reliable and precise predictions.

To demonstrate the advantage of the QIBP algorithm over the classical Bayesian approach, Table~\ref{T1} also presents the prediction results of the continuous Bayesian probability (CBP) estimator for nuclear mass \cite{k300b12, CBPRC}. The CBP estimator obtains the prior PDFs and the likelihood PDFs of the residuals through kernel density estimation (KDE). Then, it employs Bayesian theory to determine the posterior PDF and the expected value of the residuals. After optimization with the CBP estimator, the standard deviations of the WS4 and HFB models are reduced to 0.187\ MeV and 0.509\ MeV, respectively. This indicates that the predictive capability of the CBP estimator for nuclear masses is weaker than that of the QIBP algorithm.
	
	\begin{figure*}[htpb]
	\centering
	\includegraphics[width=16cm,angle=0,clip=true]{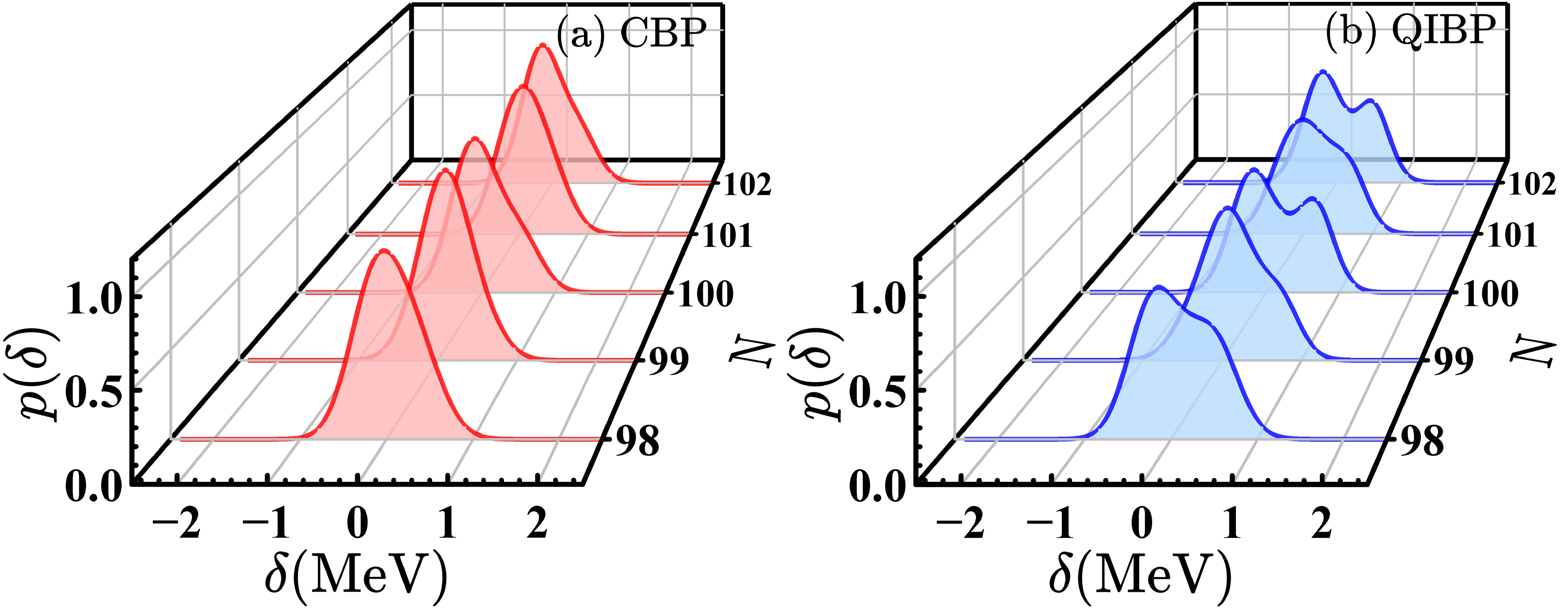}\\
	\caption{(a) Prior PDFs for Pb isotopes with $98 \le N \le 102$, obtained by the CBP estimator. (b) Prior PDFs for the same nuclei, calculated using the QIBP algorithm.}
	\label{fig2}
\end{figure*}

Moreover, we perform an in-depth comparison between the QIBP algorithm and the CBP estimator. Fig. \ref{fig2} shows the prior PDFs for Pb isotopes with $98 \le N \le 102$, obtained by the QIBP and CBP method. In Fig. \ref{fig2}(a), the CBP estimator employs Gaussian KDE to derive the prior PDFs of the mass residuals. These PDFs exhibit a single, smooth peak that primarily reflects the overall spatial distribution of the data. The construction of such PDFs is heavily influenced by adjacent nuclei. It therefore severely underestimates the impact of nuclei that are strongly correlated but not directly adjacent, such as the nuclei with the same parity of $Z$ and $N$.

By contrast, the prior PDFs calculated by Eq.~\eqref{E14} of the QIBP algorithm in Fig.~\ref{fig2}(b) exhibit a more complex structure and show a pronounced dependence on the parity of $N$. These PDFs are constructed using the Schr\"{o}dinger equation as a mathematical tool and contain the second derivatives of $\psi$, which provide enhanced sensitivity to curvature and reveal patterns that KDE cannot capture. This sensitivity is crucial for identifying structures such as shell effects and odd-even staggering. Nuclear mass residuals for nuclei with the same parity of $Z$ and $N$ or other strongly correlated nuclei tend to form clusters. The potential function identifies these clusters, yields lower potential energy, and forms potential wells. Since lower energy corresponds to higher probability density in the Boltzmann distribution, the resulting PDFs exhibit local maxima at the positions of these potential wells. These PDFs strengthen the influence of strongly correlated nuclei on the prediction rather than merely fitting the spatial distribution of data. Therefore, the PDFs employed by the QIBP algorithm are more suitable for nuclear mass prediction tasks than those of the CBP estimator.
	
\begin{figure}[h]
	\centering
	\includegraphics[width=8cm,angle=0,clip=true]{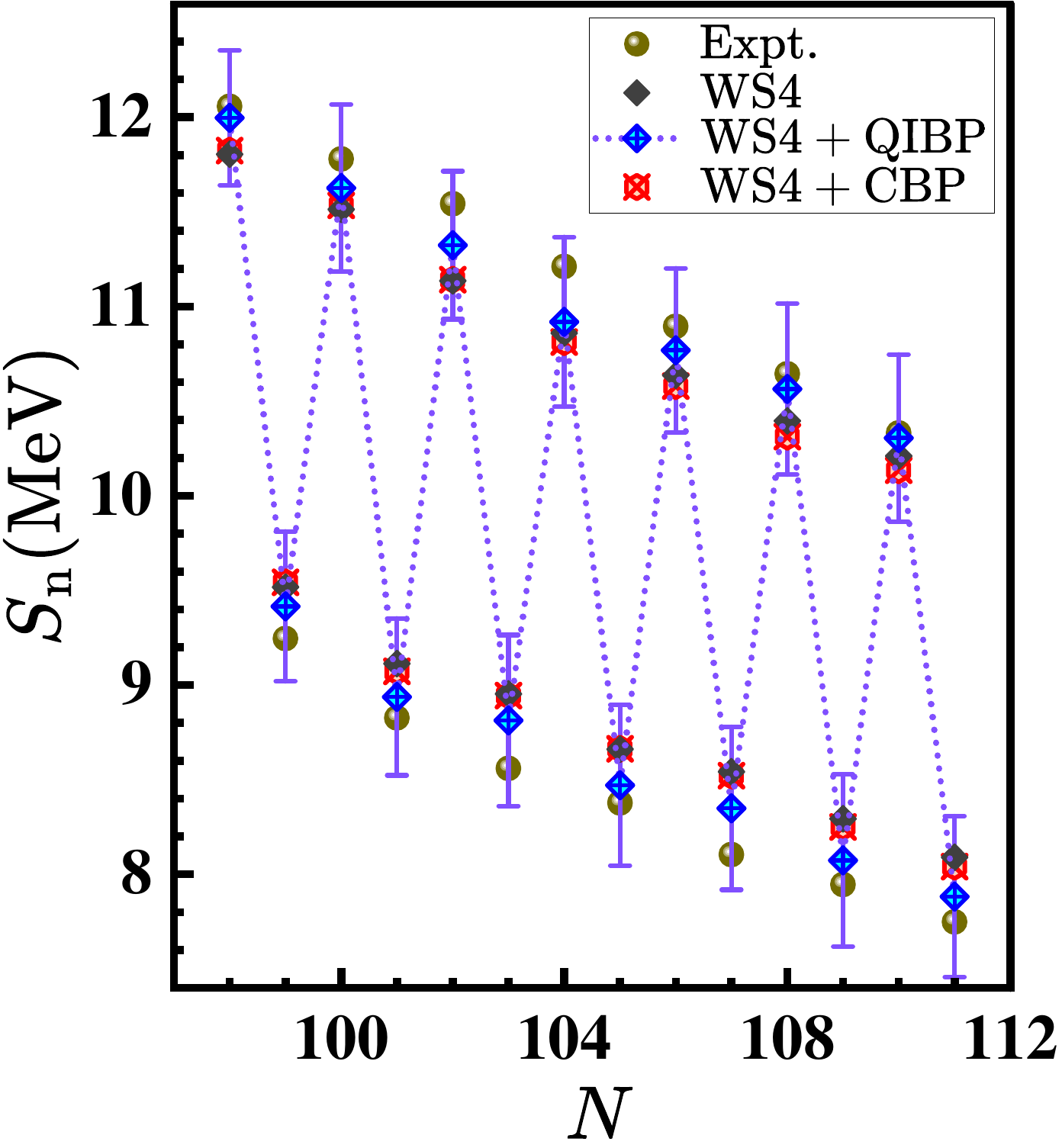}\\
	\caption{One-neutron separation energies $S_{\mathrm{n}}$ (MeV) for Pb isotopes. Brown circles denote experimental data, blue diamonds with error bars represent the $S_{\mathrm{n}}$ values and corresponding uncertainties obtained using the QIBP algorithm, and red circles and brown diamonds correspond to the results from the CBP estimator and the WS4 model, respectively.}
	\label{fig3}
\end{figure}

Extending the above discussion, the one-neutron separation energies $S_{\mathrm{n}}$ for Pb isotopes are derived from the nuclear masses obtained by the QIBP algorithm, as shown in Fig.~\ref{fig3}. The uncertainties of the QIBP algorithm shown in Fig.~\ref{fig3} are obtained from Eq.~\eqref{E20} together with the standard error-propagation formula. Fig.~\ref{fig3} also includes $S_{\mathrm{n}}$ from the WS4 model and the CBP estimator, together with experimental values for comparison. In Fig.~\ref{fig3}, the WS4 model systematically underestimates $S_{\mathrm{n}}$ for even-$N$ nuclei and overestimates $S_{\mathrm{n}}$ for odd-$N$ nuclei, indicating that part of its deviation arises from odd-even staggering. For each nucleus, the $S_{\mathrm{n}}$ values estimated by the QIBP algorithm are closer to the experimental data than the original WS4 results and those estimated by the CBP estimator. These results indicate that the PDFs obtained by the QIBP algorithm can account for complex and subtle correlations in nuclear mass data, thereby reliably capturing effects such as odd-even staggering. In contrast, the predictions of the CBP estimator are overly influenced by adjacent nuclei, which weakens its ability to observe odd-even effects. As a result, The QIBP algorithm enables more robust descriptions of nuclear masses and separation energies. In the future, we will further explore the performance and potential advantages of the QIBP algorithm compared with other classical machine learning algorithms in the field of nuclear physics.

\subsection{Extrapolating capabilities of the QIBP algorithm}
	
Evaluating the performance of the QIBP algorithm when facing unknown data is important. In this section, the extrapolation capability of the QIBP algorithm is analyzed. The training set consists of 2183 nuclear masses from AME2016 with $Z \ge 20$ and $N \ge 20$ \cite{k2}. The test set is composed of 70 newly added nuclei in AME2020. Table~\ref{T3} presents the standard deviations $\sigma^{\mathrm{pre}}$ and $\sigma^{\mathrm{post}}$ for both the training and test sets, before and after applying the QIBP algorithm, along with the improvement rates $\Delta\sigma/\sigma^{\mathrm{pre}}$.

	\begin{table}[ht]
		\caption{Raw standard deviations $\sigma_{\text{pre}}$ (MeV) from the WS4 and HFB models and the standard deviations $\sigma_{\text{post}}$ (MeV) after QIBP algorithm refinements. The training set includes 2183 nuclei with $Z\ge20$ and $N\ge20$ from AME2016, and the test set includes the newly added 70 nuclei in AME2020.}
		\label{T3}
		\centering
		\resizebox{1\linewidth}{!}{%
			\renewcommand\arraystretch{1.2}
			\begin{tabular*}{9cm}{@{\extracolsep\fill}lccccccc}
				\hline\hline
				\multirow{2}{*}{Methods} & \multirow{2}{*}{Models} & \multicolumn{3}{c}{Training set} & \multicolumn{3}{c}{Test set} \\
				\cline{3-5}\cline{6-8}
				&& $\sigma_{\text{pre}}$ & $\sigma_{\text{post}}$ & $\Delta\sigma/\sigma_{\text{pre}}$ & $\sigma_{\text{pre}}$ & $\sigma_{\text{post}}$ & $\Delta\sigma/\sigma_{\text{pre}}$ \\
				\hline
				\multirow{2}{*}{QIBP}& WS4 & 0.270 & 0.150 & 44.4\% & 0.355 & 0.216 & 39.2\% \\
				&HFB & 5.210 & 0.335 & 93.6\% & 6.367 & 0.676 & 89.4\% \\
				\hline
				\multirow{2}{*}{CBP}& WS4 & 0.270 & 0.193 & 28.5\% & 0.355 & 0.255 & 28.2\% \\
				&HFB & 5.210 & 0.526 & 89.9\% & 6.367 & 0.971 & 84.7\% \\
				
				\hline\hline
		\end{tabular*}}
	\end{table}

According to Table~\ref{T3}, for the WS4 model, although the standard deviation on the test set is slightly lower than that on the training set, it remains at a relatively high precision level of 0.355 MeV. This suggests that the theoretical model possesses a certain extrapolation capability. After refinement by the QIBP algorithm, the posterior standard deviation $\sigma^{\mathrm{post}}$ in the test set is reduced to 0.216 MeV, corresponding to an improvement rate of 39.2\%. Compared with the training set, the improvement rate decreases by only 5.2\%, demonstrating that the QIBP algorithm retains extrapolation ability when applied to the WS4 model. For the HFB model, its original results yield nearly identical values of $\sigma^{\mathrm{pre}}$ for the training and test sets, indicating a reasonable degree of extrapolation. After optimization by the QIBP algorithm, the standard deviation in the test set is reduced to 0.676 MeV, achieving an improvement rate of 89.4\%. Although $\sigma^{\mathrm{post}}$ increases compared with the training set, the improvement remains close to 90\%, and markedly enhancing the HFB model’s description of nuclear masses. Table~\ref{T3} also presents the extrapolation performance of the CBP estimator, indicating that the QIBP algorithm provides more reliable predictions for unknown nuclei than the CBP estimator.

The results demonstrate that, on the basis of the global description provided by theoretical models, the QIBP algorithm can further improve nuclear mass predictions for nuclei located at the boundary of the known region. This is due to the fact that the QIBP algorithm constructs PDFs using more effective Schr\"{o}dinger potentials to process nuclear data, enabling accurate modeling of nuclear mass residual distributions. In addition, Bayesian theory can also analyze the statistical correlations of nuclear masses. Consequently, the QIBP algorithm, constructed using the Schr\"{o}dinger equation as a mathematical tool, is capable of detecting effects not incorporated into theoretical models, such as deformation, odd-even staggering, and shell effect.

\subsection{Further extrapolation of the QIBP algorithm across the nuclear chart}

This section examines the extrapolation performance of the QIBP algorithm across the entire nuclear chart, from the boundary of the known region to the drip line. Since the experimentally known nuclear masses compiled in AME2020 are relatively limited, the performance of the QIBP algorithm in most regions far from the $\beta$-stability cannot be validated directly using experimental data. To assess its performance in all regions within the drip lines where experimental data are unavailable, the nuclear masses predicted by the WS4 model are adopted as a ``synthetic experimental set'', while the HFB model is taken as the theoretical model to be refined. The QIBP algorithm is then used to estimate the residuals between the HFB model and the WS4 model, thereby making the predictions of the HFB model closer to those of the WS4 model. The training set includes 2253 nuclei from AME2020 with $Z \ge 20$ and $N \ge 20$, while the test set comprises 2297 nuclei outside AME2020 and within the drip line determined by the HFB model, covering $20 \le Z \le 110$ and $20 \le N \le 160$.

\begin{figure*}[htpb]
		\centering
		\includegraphics[width=16cm,angle=0,clip=true]{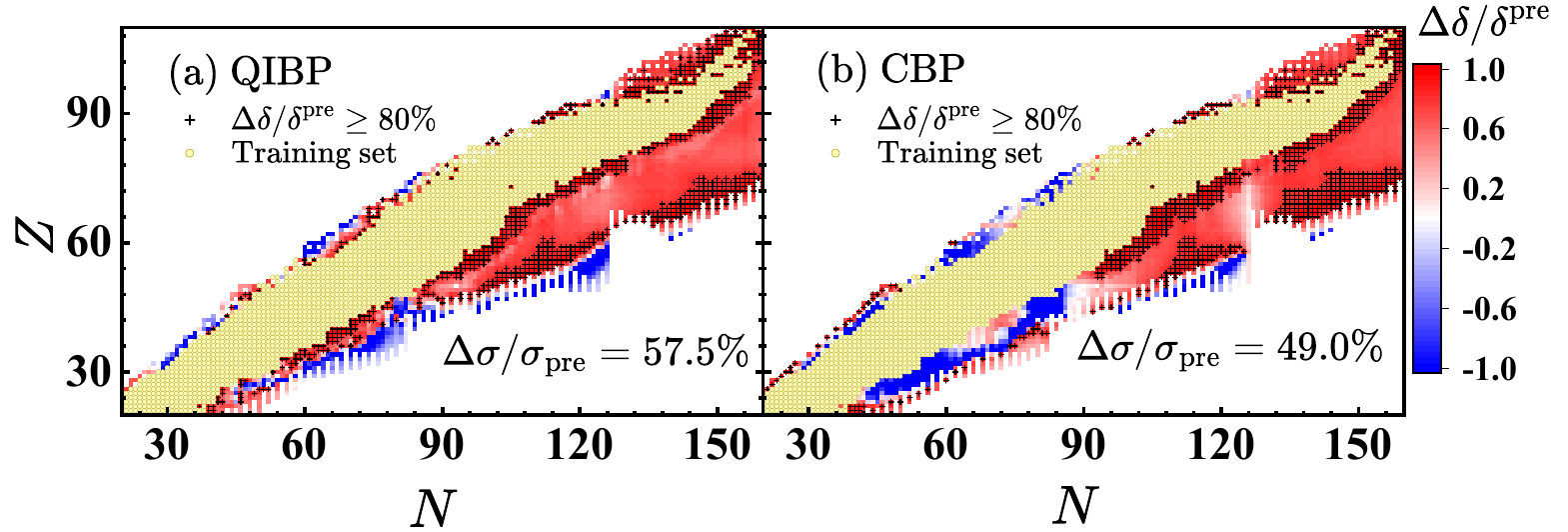}\\
		\caption{(a) Improvement ratio $\Delta\delta/\delta^{\mathrm{pre}}$ for each nucleus in the test set obtained by the QIBP algorithm, with the values mapped to colors according to the color bar on the right. The black crosses denote nuclei with $\Delta\delta/\delta^{\mathrm{pre}} \ge 80\%$. The “synthetic experimental set” corresponds to nuclear masses predicted by the WS4 model, while the theoretical model to be optimized is the HFB model. The training set includes 2253 nuclei from AME2020 with $Z \ge 20$ and $N \ge 20$, which are marked by khaki circles. The test set comprises 2297 nuclei within the drip line determined by the HFB model, covering $20 \le Z \le 110$ and $20 \le N \le 160$. (b) Same as (a), but for the CBP estimator.}
		\label{fig4}
\end{figure*}

Given that the WS4 and HFB models are based on different physical assumptions, the standard deviation between their predictions of the nuclear mass in the test set is $\sigma_{\text{pre}} = 10.660$ MeV. After applying the QIBP algorithm, the standard deviation is reduced to $\sigma_{\text{post}} = 4.532$ MeV, which represents a 57.5\% decrease compared with $\sigma_{\text{pre}}$. Fig. \ref{fig4}(a) further presents the improvement ratio $\Delta\delta/\delta^{\mathrm{pre}} = (|\delta^{\mathrm{pre}}| - |\delta^{\mathrm{post}}|)/|\delta^{\mathrm{pre}}|$ of the QIBP algorithm for each nucleus, where $\delta^{\mathrm{pre}}$ and $\delta^{\mathrm{post}}$ represent the nuclear mass residuals relative to the synthetic experimental set before and after refinement with the QIBP algorithm, respectively. In Fig. \ref{fig4}(a), the QIBP algorithm achieves broad improvement across the nuclear chart. The black crosses mark nuclei with $\Delta\delta/\delta^{\mathrm{pre}} \ge 80\%$, which are uniformly distributed around the known region, indicating that the QIBP algorithm significantly improves the theoretical models’ predictions of nuclear masses in these regions.

From the above discussion, it is clear that the QIBP algorithm can provide reliable nuclear mass predictions for nuclei near the experimentally known region. For nuclei far from the known region, although the QIBP algorithm globally adjusts HFB predictions toward the WS4 results, its ability to capture subtle physical effects in distant regions remains limited. When extrapolating toward the drip lines, several nuclei display negative values of the improvement ratio, indicating that the QIBP algorithm worsens the predictions of the original model for these nuclei. This results from the lack of nearby nuclei in the training set to provide sufficient physical information about residual variations in these regions, which prevents the QIBP algorithm from capturing finer physical effects. In addition, Eqs.~(\ref{E2}) and~(\ref{E3}) show that the predictions of the QIBP algorithm rely on nuclides in the training set with the same $Z$ or $N$ as the target nucleus. If no such nuclides are present in the training set, the QIBP algorithm is not applicable. These limitations are expected to be gradually resolved as more nuclear masses are precisely measured in experiments and as the algorithmic framework is further improved.
	
Additionally, the same validation is performed for the CBP estimator. The $\sigma^{\mathrm{post}}$ obtained by the CBP estimator is 5.441 MeV, representing a 49.0\% decrease compared with $\sigma^{\mathrm{pre}}$. This indicates that the overall performance of the CBP estimator is weaker than that of the QIBP algorithm. Fig. \ref{fig4}(b) shows the optimization results of the CBP estimator in different regions. It can be seen that the QIBP algorithm exhibits more broadly and uniformly distributed regions with $\Delta\delta/\delta^{\mathrm{pre}} \ge 80\%$ than the CBP estimator, especially in the regions with $N \le 90$. Moreover, at the edges of the known region, the QIBP algorithm exhibits fewer areas with $\Delta\delta/\delta^{\mathrm{pre}} \le 0\%$ compared with the CBP estimator. This is due to the QIBP algorithm employing a more robust potential function to process nuclear mass data. It effectively accounts for more complex correlations and patterns in nuclear mass data and thus achieves better extrapolation than the CBP estimator.

\subsection{Predictions of $\alpha$-decay energy based on the QIBP algorithm}
	
\begin{figure*}[htpb]
	\centering
	\hspace*{-9mm}\includegraphics[width=16cm,angle=0,clip=true]{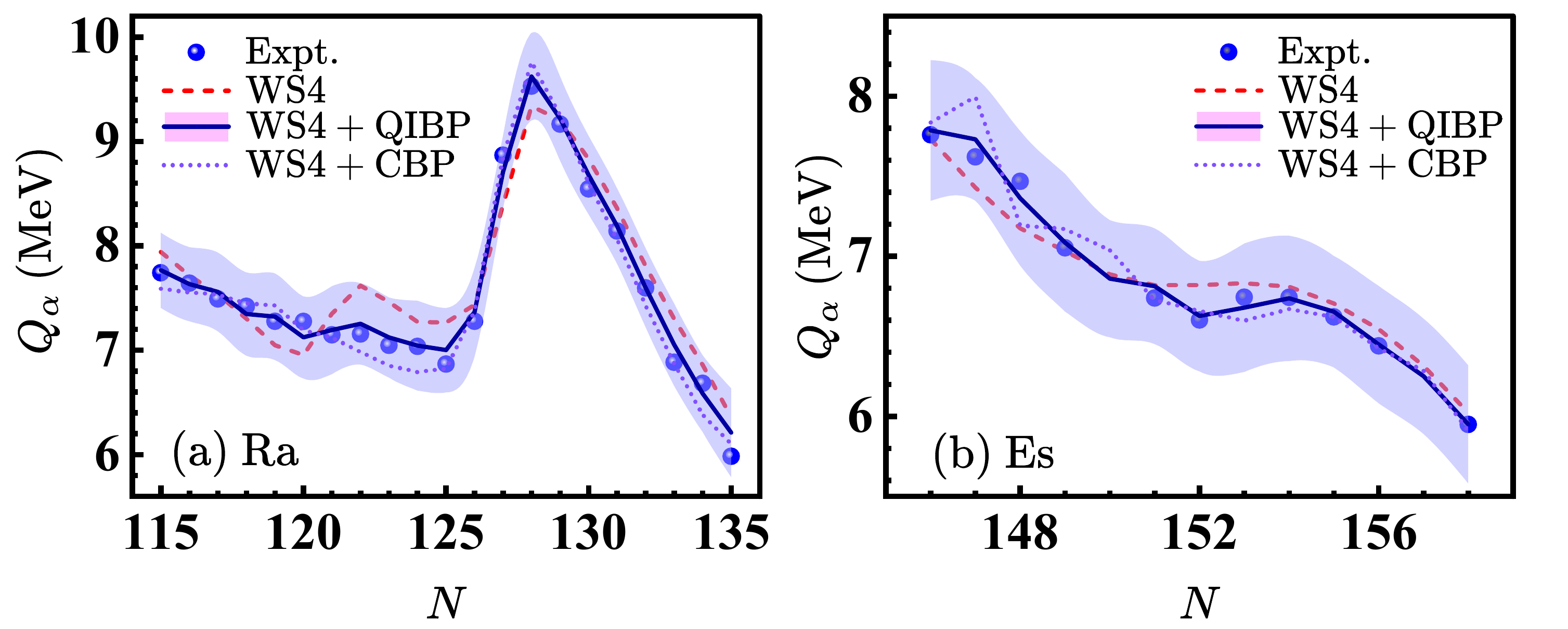}\\
	\caption{(a) $\alpha$-decay energies $Q_\alpha$ (MeV) for the Ra isotope chain. The blue spheres represent the experimental values, the red curve shows the original predictions from the WS4 model, and the blue curve shows the results after optimization by the QIBP algorithm. The blue shaded region denotes the uncertainty estimated by the QIBP algorithm. The blue dotted line denotes the $Q_\alpha$ values obtained by optimizing WS4 with the CBP estimator. (b) The same as (a), but for the Es isotope chain.}
	\label{fig5}
\end{figure*}

This section analyzes the predictive capability of the QIBP algorithm for $\alpha$-decay energy $Q_\alpha$. As a key quantity derived from nuclear mass, $Q_\alpha$ serves as an essential input for studying the $\alpha$-decay process \cite{k3, k4}. Due to the logarithmic dependency of the $\alpha$-decay half-life $T_{1/2}$ on $Q_\alpha$, as described by the Geiger-Nuttall law, even minor deviations in the predicted $Q_\alpha$ can result in significant differences in the calculated $T_{1/2}$ \cite{c1}. Furthermore, accurate $Q_\alpha$ values are indispensable for probing the properties of superheavy nuclei \cite{d1}, understanding the shell structure \cite{d2}, and identifying new nuclides \cite{d3}. $Q_\alpha$ values for the Ra and Es isotopic chains are calculated using the QIBP algorithm based on the WS4 model, as shown in Fig.~\ref{fig5}. In this study, $Q_{\alpha}$ values are derived from the nuclear masses obtained with the QIBP algorithm, thereby demonstrating the capability of QIBP to improve the WS4 model’s description of nuclear mass variations. The corresponding uncertainty is determined using Eq.~\eqref{E20} of the one-sigma uncertainty of the nuclear mass with the standard error-propagation formula. For comparison, Fig.~\ref{fig5} also includes the results of $Q_\alpha$ from CBP estimator and the original results from the WS4 model.

Fig. \ref{fig5} shows that the WS4 model successfully captures the overall trend of $Q_\alpha$ variations, but discrepancies persist in specific regions. For Ra isotopes, the WS4 model accurately reproduces the sharp increase in $Q_\alpha$ at $N=126$ associated with shell closure. However, in the neutron-deficient region, its original predictions deviate to some extent from the experimental values. For Es isotopes, many studies have verified the presence of a subshell at $N=152$ \cite{k8, k9, k10}, which is manifested as an enhancement in $Q_\alpha$. Fig. \ref{fig5}(b) indicates that the original predictions of the WS4 model do not fully account for the subshell effect in this region.

After optimization by the QIBP algorithm, the description of $Q_\alpha$ is improved for nearly all nuclei. For Ra isotopes, the QIBP algorithm applies slight adjustments to the WS4 predictions, making them more consistent with the experimental values. For Es isotopes, the improvement achieved by the QIBP algorithm is likewise evident, yielding a more accurate description of the subshell effect at $N=152$. This improvement is attributed to the application of the Schr\"{o}dinger equation as a mathematical tool during the training process. By processing nuclear data through robust Schr\"{o}dinger potentials, the QIBP algorithm effectively identifies correlations and patterns among nuclei, thereby capturing physical effects such as subshell effects, and yielding more reliable predictions.

To further demonstrate the advantage of the QIBP algorithm in identifying subtle effects and correlations, Fig.~\ref{fig5} also presents the $Q_\alpha$ predictions obtained using the CBP estimator. It can be seen that, for both Ra and Es isotopes, the $Q_{\alpha}$ values predicted by the QIBP algorithm are more consistent with the experimental values than those predicted by the CBP estimator. In particular, for the Es isotopes at $N=152$, the QIBP algorithm captures the influence of the subshell effect on $Q_\alpha$ more accurately than the CBP estimator. Since the QIBP algorithm employs a more robust Schr\"{o}dinger potential to process nuclear mass data, it can effectively account for the influences of strongly correlated nuclei. As a result, it provides a more reliable description of physical effects in nuclear mass data, such as subshell effects. In contrast, the predictions of the CBP estimator rely excessively on adjacent nuclei, which reduces the influence of strongly correlated but non-adjacent nuclei on the prediction and thus weakens its ability to capture these physical effects.

\section{Summary and outlook}\label{sec:4}
	
In this study, a novel quantum-inspired Bayesian probability (QIBP) algorithm is proposed to optimize theoretical nuclear mass models. In global optimization, the QIBP algorithm achieves improvements of 45.4\% for the WS4 model and 93.8\% for the HFB model, reducing their standard deviations to 0.149\ MeV and 0.324\,MeV, respectively. In extrapolation analysis, the improvements remain substantial at 39.2\% for the WS4 model and 89.4\% for the HFB model, confirming the extrapolation capability of the QIBP algorithm. Additionally, extrapolation using a synthetic experimental set shows that the QIBP algorithm appreciably improves the description of nuclear masses near the boundary of the known region. Furthermore, the QIBP algorithm improves the WS4 model's predictions of $\alpha$-decay energies and provides a more reliable description of shell effects. 
	
The QIBP algorithm still has limitations for nuclides far from the known region. Extrapolation based on the synthetic experimental set shows that, as the extrapolation distance increases, the stability and applicability of the QIBP algorithm gradually decrease. In particular, for some nuclei near the drip lines, it worsens the predictive performance of the original theoretical model for nuclear masses. This is mainly because known data are extremely scarce in these regions to provide the relevant physical information. Additionally, when the training set contains no nuclides with the same proton number or neutron number as the target nucleus, the QIBP algorithm is no longer applicable. These limitations are expected to be gradually alleviated as more nuclear mass data are precisely measured in experiments or as additional physical constraints are incorporated explicitly into the QIBP algorithm.
	
The effectiveness of the QIBP algorithm can be attributed to its theoretical framework. Specifically, the algorithm maps nuclear mass residuals into Hilbert space and derives probability density functions through wave functions, Schr\"{o}dinger potentials, and the Boltzmann distribution. Using the Schr\"{o}dinger equation as a mathematical tool to construct the PDFs, the QIBP algorithm can effectively capture physical effects in nuclear masses, including shell structure, odd-even staggering, and deformation. In summary, the QIBP algorithm is successfully applied to nuclear mass predictions and can provide reliable results for unknown nuclei near the experimentally known region. This algorithm also exhibits scalability and can be applied to future research in other fields such as nuclear reactions and astrophysics.

\section*{Acknowledgements}
	
This work was supported by the National Natural Science Foundation of China (Grants No. 12475135, No. 12035011 and No. 12475119), by the Shandong Provincial Natural Science Foundation, China (Grant No. ZR2020MA096), and by the Fundamental Research Funds for the Central Universities (Grant No. 22CX03017A).
	
\section*{Data availability}
	
The experimental data that support the findings of this article are openly available \cite{k1,k2}. The data that were calculated to support the findings of this article are not publicly available upon publication because it is not technically feasible and/or the cost of preparing, depositing, and hosting the data would be prohibitive within the terms of this research project. The data are available from the authors upon reasonable request.
	
\bibliography{QIBP}

\begin{thebibliography}{95}%
\makeatletter
\providecommand \@ifxundefined [1]{%
 \@ifx{#1\undefined}
}%
\providecommand \@ifnum [1]{%
 \ifnum #1\expandafter \@firstoftwo
 \else \expandafter \@secondoftwo
 \fi
}%
\providecommand \@ifx [1]{%
 \ifx #1\expandafter \@firstoftwo
 \else \expandafter \@secondoftwo
 \fi
}%
\providecommand \natexlab [1]{#1}%
\providecommand \enquote  [1]{``#1''}%
\providecommand \bibnamefont  [1]{#1}%
\providecommand \bibfnamefont [1]{#1}%
\providecommand \citenamefont [1]{#1}%
\providecommand \href@noop [0]{\@secondoftwo}%
\providecommand \href [0]{\begingroup \@sanitize@url \@href}%
\providecommand \@href[1]{\@@startlink{#1}\@@href}%
\providecommand \@@href[1]{\endgroup#1\@@endlink}%
\providecommand \@sanitize@url [0]{\catcode `\\12\catcode `\$12\catcode
  `\&12\catcode `\#12\catcode `\^12\catcode `\_12\catcode `\%12\relax}%
\providecommand \@@startlink[1]{}%
\providecommand \@@endlink[0]{}%
\providecommand \url  [0]{\begingroup\@sanitize@url \@url }%
\providecommand \@url [1]{\endgroup\@href {#1}{\urlprefix }}%
\providecommand \urlprefix  [0]{URL }%
\providecommand \Eprint [0]{\href }%
\providecommand \doibase [0]{https://doi.org/}%
\providecommand \selectlanguage [0]{\@gobble}%
\providecommand \bibinfo  [0]{\@secondoftwo}%
\providecommand \bibfield  [0]{\@secondoftwo}%
\providecommand \translation [1]{[#1]}%
\providecommand \BibitemOpen [0]{}%
\providecommand \bibitemStop [0]{}%
\providecommand \bibitemNoStop [0]{.\EOS\space}%
\providecommand \EOS [0]{\spacefactor3000\relax}%
\providecommand \BibitemShut  [1]{\csname bibitem#1\endcsname}%
\let\auto@bib@innerbib\@empty
%</preamble>
\bibitem [{\citenamefont {Jordan}\ and\ \citenamefont
  {Mitchell}(2015)}]{k1001}%
  \BibitemOpen
  \bibfield  {author} {\bibinfo {author} {\bibfnamefont {M.~I.}\ \bibnamefont
  {Jordan}}\ and\ \bibinfo {author} {\bibfnamefont {T.~M.}\ \bibnamefont
  {Mitchell}},\ }\href {https://doi.org/10.1126/science.aaa8415} {\bibfield
  {journal} {\bibinfo  {journal} {Science}\ }\textbf {\bibinfo {volume}
  {349}},\ \bibinfo {pages} {255} (\bibinfo {year} {2015})}\BibitemShut
  {NoStop}%
\bibitem [{\citenamefont {Carleo}\ \emph {et~al.}(2019)\citenamefont {Carleo},
  \citenamefont {Cirac}, \citenamefont {Cranmer}, \citenamefont {Daudet},
  \citenamefont {Schuld}, \citenamefont {Tishby}, \citenamefont
  {Vogt-Maranto},\ and\ \citenamefont {Zdeborov\'a}}]{k1002}%
  \BibitemOpen
  \bibfield  {author} {\bibinfo {author} {\bibfnamefont {G.}~\bibnamefont
  {Carleo}}, \bibinfo {author} {\bibfnamefont {I.}~\bibnamefont {Cirac}},
  \bibinfo {author} {\bibfnamefont {K.}~\bibnamefont {Cranmer}}, \bibinfo
  {author} {\bibfnamefont {L.}~\bibnamefont {Daudet}}, \bibinfo {author}
  {\bibfnamefont {M.}~\bibnamefont {Schuld}}, \bibinfo {author} {\bibfnamefont
  {N.}~\bibnamefont {Tishby}}, \bibinfo {author} {\bibfnamefont
  {L.}~\bibnamefont {Vogt-Maranto}},\ and\ \bibinfo {author} {\bibfnamefont
  {L.}~\bibnamefont {Zdeborov\'a}},\ }\href
  {https://doi.org/10.1103/RevModPhys.91.045002} {\bibfield  {journal}
  {\bibinfo  {journal} {Rev. Mod. Phys.}\ }\textbf {\bibinfo {volume} {91}},\
  \bibinfo {pages} {045002} (\bibinfo {year} {2019})}\BibitemShut {NoStop}%
\bibitem [{\citenamefont {Butler}\ \emph {et~al.}(2018)\citenamefont {Butler},
  \citenamefont {Davies}, \citenamefont {Cartwright}, \citenamefont {Isayev},\
  and\ \citenamefont {Walsh}}]{1.e1}%
  \BibitemOpen
  \bibfield  {author} {\bibinfo {author} {\bibfnamefont {K.~T.}\ \bibnamefont
  {Butler}}, \bibinfo {author} {\bibfnamefont {D.~W.}\ \bibnamefont {Davies}},
  \bibinfo {author} {\bibfnamefont {H.}~\bibnamefont {Cartwright}}, \bibinfo
  {author} {\bibfnamefont {O.}~\bibnamefont {Isayev}},\ and\ \bibinfo {author}
  {\bibfnamefont {A.}~\bibnamefont {Walsh}},\ }\href
  {https://doi.org/10.1038/s41586-018-0337-2} {\bibfield  {journal} {\bibinfo
  {journal} {Nature}\ }\textbf {\bibinfo {volume} {559}},\ \bibinfo {pages}
  {547} (\bibinfo {year} {2018})}\BibitemShut {NoStop}%
\bibitem [{\citenamefont {Greener}\ \emph {et~al.}(2022)\citenamefont
  {Greener}, \citenamefont {Kandathil}, \citenamefont {Moffat},\ and\
  \citenamefont {Jones}}]{1.e2}%
  \BibitemOpen
  \bibfield  {author} {\bibinfo {author} {\bibfnamefont {J.~G.}\ \bibnamefont
  {Greener}}, \bibinfo {author} {\bibfnamefont {S.~M.}\ \bibnamefont
  {Kandathil}}, \bibinfo {author} {\bibfnamefont {L.}~\bibnamefont {Moffat}},\
  and\ \bibinfo {author} {\bibfnamefont {D.~T.}\ \bibnamefont {Jones}},\ }\href
  {https://doi.org/10.1038/s41580-021-00407-0} {\bibfield  {journal} {\bibinfo
  {journal} {Nat. Rev. Mol. Cell Biol.}\ }\textbf {\bibinfo {volume} {23}},\
  \bibinfo {pages} {40} (\bibinfo {year} {2022})}\BibitemShut {NoStop}%
\bibitem [{1.d({\natexlab{a}})}]{1.d1}%
  \BibitemOpen
  https://www.nobelprize.org/prizes/physics/2024\BibitemShut {NoStop}%
\bibitem [{\citenamefont {Hopfield}(1982)}]{k1005}%
  \BibitemOpen
  \bibfield  {author} {\bibinfo {author} {\bibfnamefont {J.~J.}\ \bibnamefont
  {Hopfield}},\ }\href {https://doi.org/10.1073/pnas.79.8.2554} {\bibfield
  {journal} {\bibinfo  {journal} {Proc. Natl. Acad. Sci.}\ }\textbf {\bibinfo
  {volume} {79}},\ \bibinfo {pages} {2554} (\bibinfo {year}
  {1982})}\BibitemShut {NoStop}%
\bibitem [{\citenamefont {Rumelhart}\ \emph {et~al.}(1986)\citenamefont
  {Rumelhart}, \citenamefont {Hinton},\ and\ \citenamefont {Williams}}]{k1006}%
  \BibitemOpen
  \bibfield  {author} {\bibinfo {author} {\bibfnamefont {D.~E.}\ \bibnamefont
  {Rumelhart}}, \bibinfo {author} {\bibfnamefont {G.~E.}\ \bibnamefont
  {Hinton}},\ and\ \bibinfo {author} {\bibfnamefont {R.~J.}\ \bibnamefont
  {Williams}},\ }\href {https://doi.org/10.1038/323533a0} {\bibfield  {journal}
  {\bibinfo  {journal} {Nature}\ }\textbf {\bibinfo {volume} {323}},\ \bibinfo
  {pages} {533} (\bibinfo {year} {1986})}\BibitemShut {NoStop}%
\bibitem [{\citenamefont {Hinton}(2025)}]{RMP1}%
  \BibitemOpen
  \bibfield  {author} {\bibinfo {author} {\bibfnamefont {G.}~\bibnamefont
  {Hinton}},\ }\href {https://doi.org/10.1103/RevModPhys.97.030502} {\bibfield
  {journal} {\bibinfo  {journal} {Rev. Mod. Phys.}\ }\textbf {\bibinfo {volume}
  {97}},\ \bibinfo {pages} {030502} (\bibinfo {year} {2025})}\BibitemShut
  {NoStop}%
\bibitem [{\citenamefont {Hopfield}(2025)}]{RMP2}%
  \BibitemOpen
  \bibfield  {author} {\bibinfo {author} {\bibfnamefont {J.~J.}\ \bibnamefont
  {Hopfield}},\ }\href {https://doi.org/10.1103/RevModPhys.97.030501}
  {\bibfield  {journal} {\bibinfo  {journal} {Rev. Mod. Phys.}\ }\textbf
  {\bibinfo {volume} {97}},\ \bibinfo {pages} {030501} (\bibinfo {year}
  {2025})}\BibitemShut {NoStop}%
\bibitem [{1.d({\natexlab{b}})}]{1.d2}%
  \BibitemOpen
  https://www.nobelprize.org/prizes/chemistry/2024\BibitemShut {NoStop}%
\bibitem [{\citenamefont {Senior}\ \emph {et~al.}(2020)\citenamefont {Senior},
  \citenamefont {Evans}, \citenamefont {Jumper}, \citenamefont {Kirkpatrick},
  \citenamefont {Sifre}, \citenamefont {Green}, \citenamefont {Qin},
  \citenamefont {{\v{Z}}{\'\i}dek}, \citenamefont {Nelson}, \citenamefont
  {Bridgland}, \citenamefont {Penedones}, \citenamefont {Petersen},
  \citenamefont {Simonyan}, \citenamefont {Crossan}, \citenamefont {Kohli},
  \citenamefont {Jones}, \citenamefont {Silver}, \citenamefont {Kavukcuoglu},\
  and\ \citenamefont {Hassabis}}]{k1007}%
  \BibitemOpen
  \bibfield  {author} {\bibinfo {author} {\bibfnamefont {A.~W.}\ \bibnamefont
  {Senior}}, \bibinfo {author} {\bibfnamefont {R.}~\bibnamefont {Evans}},
  \bibinfo {author} {\bibfnamefont {J.}~\bibnamefont {Jumper}}, \bibinfo
  {author} {\bibfnamefont {J.}~\bibnamefont {Kirkpatrick}}, \bibinfo {author}
  {\bibfnamefont {L.}~\bibnamefont {Sifre}}, \bibinfo {author} {\bibfnamefont
  {T.}~\bibnamefont {Green}}, \bibinfo {author} {\bibfnamefont
  {C.}~\bibnamefont {Qin}}, \bibinfo {author} {\bibfnamefont {A.}~\bibnamefont
  {{\v{Z}}{\'\i}dek}}, \bibinfo {author} {\bibfnamefont {A.~W.~R.}\
  \bibnamefont {Nelson}}, \bibinfo {author} {\bibfnamefont {A.}~\bibnamefont
  {Bridgland}}, \emph {et~al.},\ }\href
  {https://doi.org/10.1038/s41586-019-1923-7} {\bibfield  {journal} {\bibinfo
  {journal} {Nature}\ }\textbf {\bibinfo {volume} {577}},\ \bibinfo {pages}
  {706} (\bibinfo {year} {2020})}\BibitemShut {NoStop}%
\bibitem [{\citenamefont {Baek}\ \emph {et~al.}(2021)\citenamefont {Baek},
  \citenamefont {DiMaio}, \citenamefont {Anishchenko}, \citenamefont
  {Dauparas}, \citenamefont {Ovchinnikov}, \citenamefont {Lee}, \citenamefont
  {Wang}, \citenamefont {Cong}, \citenamefont {Kinch}, \citenamefont
  {Schaeffer}, \citenamefont {Millán}, \citenamefont {Park}, \citenamefont
  {Adams}, \citenamefont {Glassman}, \citenamefont {DeGiovanni}, \citenamefont
  {Pereira}, \citenamefont {Rodrigues}, \citenamefont {van Dijk}, \citenamefont
  {Ebrecht}, \citenamefont {Opperman}, \citenamefont {Sagmeister},
  \citenamefont {Buhlheller}, \citenamefont {Pavkov-Keller}, \citenamefont
  {Rathinaswamy}, \citenamefont {Dalwadi}, \citenamefont {Yip}, \citenamefont
  {Burke}, \citenamefont {Garcia}, \citenamefont {Grishin}, \citenamefont
  {Adams}, \citenamefont {Read},\ and\ \citenamefont {Baker}}]{k1008}%
  \BibitemOpen
  \bibfield  {author} {\bibinfo {author} {\bibfnamefont {M.}~\bibnamefont
  {Baek}}, \bibinfo {author} {\bibfnamefont {F.}~\bibnamefont {DiMaio}},
  \bibinfo {author} {\bibfnamefont {I.}~\bibnamefont {Anishchenko}}, \bibinfo
  {author} {\bibfnamefont {J.}~\bibnamefont {Dauparas}}, \bibinfo {author}
  {\bibfnamefont {S.}~\bibnamefont {Ovchinnikov}}, \bibinfo {author}
  {\bibfnamefont {G.~R.}\ \bibnamefont {Lee}}, \bibinfo {author} {\bibfnamefont
  {J.}~\bibnamefont {Wang}}, \bibinfo {author} {\bibfnamefont {Q.}~\bibnamefont
  {Cong}}, \bibinfo {author} {\bibfnamefont {L.~N.}\ \bibnamefont {Kinch}},
  \bibinfo {author} {\bibfnamefont {R.~D.}\ \bibnamefont {Schaeffer}}, \emph
  {et~al.},\ }\href {https://doi.org/10.1126/science.abj8754} {\bibfield
  {journal} {\bibinfo  {journal} {Science}\ }\textbf {\bibinfo {volume}
  {373}},\ \bibinfo {pages} {871} (\bibinfo {year} {2021})}\BibitemShut
  {NoStop}%
\bibitem [{\citenamefont {Jumper}\ \emph {et~al.}(2021)\citenamefont {Jumper},
  \citenamefont {Evans}, \citenamefont {Pritzel}, \citenamefont {Green},
  \citenamefont {Figurnov}, \citenamefont {Ronneberger}, \citenamefont
  {Tunyasuvunakool}, \citenamefont {Bates}, \citenamefont {{\v{Z}}{\'\i}dek},
  \citenamefont {Potapenko}, \citenamefont {Bridgland}, \citenamefont {Meyer},
  \citenamefont {Kohl}, \citenamefont {Ballard}, \citenamefont {Cowie},
  \citenamefont {Romera-Paredes}, \citenamefont {Nikolov}, \citenamefont
  {Jain}, \citenamefont {Adler}, \citenamefont {Back}, \citenamefont
  {Petersen}, \citenamefont {Reiman}, \citenamefont {Clancy}, \citenamefont
  {Zielinski}, \citenamefont {Steinegger}, \citenamefont {Pacholska},
  \citenamefont {Berghammer}, \citenamefont {Bodenstein}, \citenamefont
  {Silver}, \citenamefont {Vinyals}, \citenamefont {Senior}, \citenamefont
  {Kavukcuoglu}, \citenamefont {Kohli},\ and\ \citenamefont
  {Hassabis}}]{k1009}%
  \BibitemOpen
  \bibfield  {author} {\bibinfo {author} {\bibfnamefont {J.}~\bibnamefont
  {Jumper}}, \bibinfo {author} {\bibfnamefont {R.}~\bibnamefont {Evans}},
  \bibinfo {author} {\bibfnamefont {A.}~\bibnamefont {Pritzel}}, \bibinfo
  {author} {\bibfnamefont {T.}~\bibnamefont {Green}}, \bibinfo {author}
  {\bibfnamefont {M.}~\bibnamefont {Figurnov}}, \bibinfo {author}
  {\bibfnamefont {O.}~\bibnamefont {Ronneberger}}, \bibinfo {author}
  {\bibfnamefont {K.}~\bibnamefont {Tunyasuvunakool}}, \bibinfo {author}
  {\bibfnamefont {R.}~\bibnamefont {Bates}}, \bibinfo {author} {\bibfnamefont
  {A.}~\bibnamefont {{\v{Z}}{\'\i}dek}}, \bibinfo {author} {\bibfnamefont
  {A.}~\bibnamefont {Potapenko}}, \emph {et~al.},\ }\href
  {https://doi.org/10.1038/s41586-021-03819-2} {\bibfield  {journal} {\bibinfo
  {journal} {Nature}\ }\textbf {\bibinfo {volume} {596}},\ \bibinfo {pages}
  {583} (\bibinfo {year} {2021})}\BibitemShut {NoStop}%
\bibitem [{\citenamefont {Bennett}\ and\ \citenamefont
  {DiVincenzo}(2000)}]{k100b1}%
  \BibitemOpen
  \bibfield  {author} {\bibinfo {author} {\bibfnamefont {C.~H.}\ \bibnamefont
  {Bennett}}\ and\ \bibinfo {author} {\bibfnamefont {D.~P.}\ \bibnamefont
  {DiVincenzo}},\ }\href {https://doi.org/10.1038/35005001} {\bibfield
  {journal} {\bibinfo  {journal} {Nature}\ }\textbf {\bibinfo {volume} {404}},\
  \bibinfo {pages} {247} (\bibinfo {year} {2000})}\BibitemShut {NoStop}%
\bibitem [{\citenamefont {{Deutsch}}(1985)}]{k100b2}%
  \BibitemOpen
  \bibfield  {author} {\bibinfo {author} {\bibfnamefont {D.}~\bibnamefont
  {{Deutsch}}},\ }\href {https://doi.org/10.1098/rspa.1985.0070} {\bibfield
  {journal} {\bibinfo  {journal} {Proc. R. Soc. Lond. A}\ }\textbf {\bibinfo
  {volume} {400}},\ \bibinfo {pages} {97} (\bibinfo {year} {1985})}\BibitemShut
  {NoStop}%
\bibitem [{\citenamefont {Grover}(1997)}]{1.e3}%
  \BibitemOpen
  \bibfield  {author} {\bibinfo {author} {\bibfnamefont {L.~K.}\ \bibnamefont
  {Grover}},\ }\href {https://doi.org/10.1103/PhysRevLett.79.325} {\bibfield
  {journal} {\bibinfo  {journal} {Phys. Rev. Lett.}\ }\textbf {\bibinfo
  {volume} {79}},\ \bibinfo {pages} {325} (\bibinfo {year} {1997})}\BibitemShut
  {NoStop}%
\bibitem [{\citenamefont {Havl{\'\i}{\v{c}}ek}\ \emph
  {et~al.}(2019)\citenamefont {Havl{\'\i}{\v{c}}ek}, \citenamefont
  {C{\'o}rcoles}, \citenamefont {Temme}, \citenamefont {Harrow}, \citenamefont
  {Kandala}, \citenamefont {Chow},\ and\ \citenamefont {Gambetta}}]{1.e4}%
  \BibitemOpen
  \bibfield  {author} {\bibinfo {author} {\bibfnamefont {V.}~\bibnamefont
  {Havl{\'\i}{\v{c}}ek}}, \bibinfo {author} {\bibfnamefont {A.~D.}\
  \bibnamefont {C{\'o}rcoles}}, \bibinfo {author} {\bibfnamefont
  {K.}~\bibnamefont {Temme}}, \bibinfo {author} {\bibfnamefont {A.~W.}\
  \bibnamefont {Harrow}}, \bibinfo {author} {\bibfnamefont {A.}~\bibnamefont
  {Kandala}}, \bibinfo {author} {\bibfnamefont {J.~M.}\ \bibnamefont {Chow}},\
  and\ \bibinfo {author} {\bibfnamefont {J.~M.}\ \bibnamefont {Gambetta}},\
  }\href {https://doi.org/10.1038/s41586-019-0980-2} {\bibfield  {journal}
  {\bibinfo  {journal} {Nature}\ }\textbf {\bibinfo {volume} {567}},\ \bibinfo
  {pages} {209} (\bibinfo {year} {2019})}\BibitemShut {NoStop}%
\bibitem [{\citenamefont {Das}\ and\ \citenamefont {Chakrabarti}(2008)}]{1.e5}%
  \BibitemOpen
  \bibfield  {author} {\bibinfo {author} {\bibfnamefont {A.}~\bibnamefont
  {Das}}\ and\ \bibinfo {author} {\bibfnamefont {B.~K.}\ \bibnamefont
  {Chakrabarti}},\ }\href {https://doi.org/10.1103/RevModPhys.80.1061}
  {\bibfield  {journal} {\bibinfo  {journal} {Rev. Mod. Phys.}\ }\textbf
  {\bibinfo {volume} {80}},\ \bibinfo {pages} {1061} (\bibinfo {year}
  {2008})}\BibitemShut {NoStop}%
\bibitem [{\citenamefont {Biamonte}\ \emph {et~al.}(2017)\citenamefont
  {Biamonte}, \citenamefont {Wittek}, \citenamefont {Pancotti}, \citenamefont
  {Rebentrost}, \citenamefont {Wiebe},\ and\ \citenamefont {Lloyd}}]{k10011}%
  \BibitemOpen
  \bibfield  {author} {\bibinfo {author} {\bibfnamefont {J.}~\bibnamefont
  {Biamonte}}, \bibinfo {author} {\bibfnamefont {P.}~\bibnamefont {Wittek}},
  \bibinfo {author} {\bibfnamefont {N.}~\bibnamefont {Pancotti}}, \bibinfo
  {author} {\bibfnamefont {P.}~\bibnamefont {Rebentrost}}, \bibinfo {author}
  {\bibfnamefont {N.}~\bibnamefont {Wiebe}},\ and\ \bibinfo {author}
  {\bibfnamefont {S.}~\bibnamefont {Lloyd}},\ }\href
  {https://doi.org/10.1038/nature23474} {\bibfield  {journal} {\bibinfo
  {journal} {Nature}\ }\textbf {\bibinfo {volume} {549}},\ \bibinfo {pages}
  {195} (\bibinfo {year} {2017})}\BibitemShut {NoStop}%
\bibitem [{\citenamefont {Cerezo}\ \emph {et~al.}(2022)\citenamefont {Cerezo},
  \citenamefont {Verdon}, \citenamefont {Huang}, \citenamefont {Cincio},\ and\
  \citenamefont {Coles}}]{k10012}%
  \BibitemOpen
  \bibfield  {author} {\bibinfo {author} {\bibfnamefont {M.}~\bibnamefont
  {Cerezo}}, \bibinfo {author} {\bibfnamefont {G.}~\bibnamefont {Verdon}},
  \bibinfo {author} {\bibfnamefont {H.-Y.}\ \bibnamefont {Huang}}, \bibinfo
  {author} {\bibfnamefont {L.}~\bibnamefont {Cincio}},\ and\ \bibinfo {author}
  {\bibfnamefont {P.~J.}\ \bibnamefont {Coles}},\ }\href
  {https://doi.org/10.1038/s43588-022-00311-3} {\bibfield  {journal} {\bibinfo
  {journal} {Nat. Comput. Sci.}\ }\textbf {\bibinfo {volume} {2}},\ \bibinfo
  {pages} {567} (\bibinfo {year} {2022})}\BibitemShut {NoStop}%
\bibitem [{\citenamefont {Sajjan}\ \emph {et~al.}(2022)\citenamefont {Sajjan},
  \citenamefont {Li}, \citenamefont {Selvarajan}, \citenamefont {Sureshbabu},
  \citenamefont {Kale}, \citenamefont {Gupta}, \citenamefont {Singh},\ and\
  \citenamefont {Kais}}]{k10015}%
  \BibitemOpen
  \bibfield  {author} {\bibinfo {author} {\bibfnamefont {M.}~\bibnamefont
  {Sajjan}}, \bibinfo {author} {\bibfnamefont {J.}~\bibnamefont {Li}}, \bibinfo
  {author} {\bibfnamefont {R.}~\bibnamefont {Selvarajan}}, \bibinfo {author}
  {\bibfnamefont {S.~H.}\ \bibnamefont {Sureshbabu}}, \bibinfo {author}
  {\bibfnamefont {S.~S.}\ \bibnamefont {Kale}}, \bibinfo {author}
  {\bibfnamefont {R.}~\bibnamefont {Gupta}}, \bibinfo {author} {\bibfnamefont
  {V.}~\bibnamefont {Singh}},\ and\ \bibinfo {author} {\bibfnamefont
  {S.}~\bibnamefont {Kais}},\ }\href {https://doi.org/10.1039/D2CS00203E}
  {\bibfield  {journal} {\bibinfo  {journal} {Chem. Soc. Rev.}\ }\textbf
  {\bibinfo {volume} {51}},\ \bibinfo {pages} {6475} (\bibinfo {year}
  {2022})}\BibitemShut {NoStop}%
\bibitem [{\citenamefont {Sch{\"u}tt}\ \emph {et~al.}(2019)\citenamefont
  {Sch{\"u}tt}, \citenamefont {Gastegger}, \citenamefont {Tkatchenko},
  \citenamefont {M{\"u}ller},\ and\ \citenamefont {Maurer}}]{k10016}%
  \BibitemOpen
  \bibfield  {author} {\bibinfo {author} {\bibfnamefont {K.~T.}\ \bibnamefont
  {Sch{\"u}tt}}, \bibinfo {author} {\bibfnamefont {M.}~\bibnamefont
  {Gastegger}}, \bibinfo {author} {\bibfnamefont {A.}~\bibnamefont
  {Tkatchenko}}, \bibinfo {author} {\bibfnamefont {K.-R.}\ \bibnamefont
  {M{\"u}ller}},\ and\ \bibinfo {author} {\bibfnamefont {R.~J.}\ \bibnamefont
  {Maurer}},\ }\href {https://doi.org/10.1038/s41467-019-12875-2} {\bibfield
  {journal} {\bibinfo  {journal} {Nat. Commun.}\ }\textbf {\bibinfo {volume}
  {10}},\ \bibinfo {pages} {5024} (\bibinfo {year} {2019})}\BibitemShut
  {NoStop}%
\bibitem [{\citenamefont {Guan}\ \emph {et~al.}(2021)\citenamefont {Guan},
  \citenamefont {Perdue}, \citenamefont {Pesah}, \citenamefont {Schuld},
  \citenamefont {Terashi}, \citenamefont {Vallecorsa},\ and\ \citenamefont
  {Vlimant}}]{k10017}%
  \BibitemOpen
  \bibfield  {author} {\bibinfo {author} {\bibfnamefont {W.}~\bibnamefont
  {Guan}}, \bibinfo {author} {\bibfnamefont {G.}~\bibnamefont {Perdue}},
  \bibinfo {author} {\bibfnamefont {A.}~\bibnamefont {Pesah}}, \bibinfo
  {author} {\bibfnamefont {M.}~\bibnamefont {Schuld}}, \bibinfo {author}
  {\bibfnamefont {K.}~\bibnamefont {Terashi}}, \bibinfo {author} {\bibfnamefont
  {S.}~\bibnamefont {Vallecorsa}},\ and\ \bibinfo {author} {\bibfnamefont
  {J.~R.}\ \bibnamefont {Vlimant}},\ }\href
  {https://doi.org/10.1088/2632-2153/abc17d} {\bibfield  {journal} {\bibinfo
  {journal} {Mach. learn. sci. technol.}\ }\textbf {\bibinfo {volume} {2}},\
  \bibinfo {pages} {011003} (\bibinfo {year} {2021})}\BibitemShut {NoStop}%
\bibitem [{\citenamefont {Blance}\ and\ \citenamefont
  {Spannowsky}(2021)}]{k10018}%
  \BibitemOpen
  \bibfield  {author} {\bibinfo {author} {\bibfnamefont {A.}~\bibnamefont
  {Blance}}\ and\ \bibinfo {author} {\bibfnamefont {M.}~\bibnamefont
  {Spannowsky}},\ }\href {https://doi.org/10.1007/JHEP02(2021)212} {\bibfield
  {journal} {\bibinfo  {journal} {J. High Energy Phys.}\ }\textbf {\bibinfo
  {volume} {2021}}\bibinfo  {number} { (2)},\ \bibinfo {pages}
  {212}}\BibitemShut {NoStop}%
\bibitem [{\citenamefont {Chen}\ \emph {et~al.}(2022)\citenamefont {Chen},
  \citenamefont {Wei}, \citenamefont {Zhang}, \citenamefont {Yu},\ and\
  \citenamefont {Yoo}}]{k10019}%
  \BibitemOpen
\bibfield  {number} {  }\bibfield  {author} {\bibinfo {author} {\bibfnamefont
  {S.~Y.~C.}\ \bibnamefont {Chen}}, \bibinfo {author} {\bibfnamefont {T.~C.}\
  \bibnamefont {Wei}}, \bibinfo {author} {\bibfnamefont {C.}~\bibnamefont
  {Zhang}}, \bibinfo {author} {\bibfnamefont {H.}~\bibnamefont {Yu}},\ and\
  \bibinfo {author} {\bibfnamefont {S.}~\bibnamefont {Yoo}},\ }\href
  {https://doi.org/10.1103/PhysRevResearch.4.013231} {\bibfield  {journal}
  {\bibinfo  {journal} {Phys. Rev. Res.}\ }\textbf {\bibinfo {volume} {4}},\
  \bibinfo {pages} {013231} (\bibinfo {year} {2022})}\BibitemShut {NoStop}%
\bibitem [{\citenamefont {Duckett}\ \emph {et~al.}(2024)\citenamefont
  {Duckett}, \citenamefont {Facini}, \citenamefont {Jastrzebski}, \citenamefont
  {Malik}, \citenamefont {Scanlon},\ and\ \citenamefont {Rettie}}]{k10020}%
  \BibitemOpen
  \bibfield  {author} {\bibinfo {author} {\bibfnamefont {P.}~\bibnamefont
  {Duckett}}, \bibinfo {author} {\bibfnamefont {G.}~\bibnamefont {Facini}},
  \bibinfo {author} {\bibfnamefont {M.}~\bibnamefont {Jastrzebski}}, \bibinfo
  {author} {\bibfnamefont {S.}~\bibnamefont {Malik}}, \bibinfo {author}
  {\bibfnamefont {T.}~\bibnamefont {Scanlon}},\ and\ \bibinfo {author}
  {\bibfnamefont {S.}~\bibnamefont {Rettie}},\ }\href
  {https://doi.org/10.1103/PhysRevD.109.052002} {\bibfield  {journal} {\bibinfo
   {journal} {Phys. Rev. D}\ }\textbf {\bibinfo {volume} {109}},\ \bibinfo
  {pages} {052002} (\bibinfo {year} {2024})}\BibitemShut {NoStop}%
\bibitem [{\citenamefont {Chen}\ and\ \citenamefont {Chen}(2025)}]{k10021}%
  \BibitemOpen
  \bibfield  {author} {\bibinfo {author} {\bibfnamefont {Y.~A.}\ \bibnamefont
  {Chen}}\ and\ \bibinfo {author} {\bibfnamefont {K.~F.}\ \bibnamefont
  {Chen}},\ }\href {https://doi.org/10.1103/PhysRevD.111.016020} {\bibfield
  {journal} {\bibinfo  {journal} {Phys. Rev. D}\ }\textbf {\bibinfo {volume}
  {111}},\ \bibinfo {pages} {016020} (\bibinfo {year} {2025})}\BibitemShut
  {NoStop}%
\bibitem [{\citenamefont {Martín-Guerrero}\ and\ \citenamefont
  {Lamata}(2022)}]{k10010}%
  \BibitemOpen
  \bibfield  {author} {\bibinfo {author} {\bibfnamefont {J.~D.}\ \bibnamefont
  {Martín-Guerrero}}\ and\ \bibinfo {author} {\bibfnamefont {L.}~\bibnamefont
  {Lamata}},\ }\href
  {https://doi.org/https://doi.org/10.1016/j.neucom.2021.02.102} {\bibfield
  {journal} {\bibinfo  {journal} {Neurocomputing}\ }\textbf {\bibinfo {volume}
  {470}},\ \bibinfo {pages} {457} (\bibinfo {year} {2022})}\BibitemShut
  {NoStop}%
\bibitem [{\citenamefont {Lloyd}\ \emph {et~al.}(2014)\citenamefont {Lloyd},
  \citenamefont {Mohseni},\ and\ \citenamefont {Rebentrost}}]{k2004}%
  \BibitemOpen
  \bibfield  {author} {\bibinfo {author} {\bibfnamefont {S.}~\bibnamefont
  {Lloyd}}, \bibinfo {author} {\bibfnamefont {M.}~\bibnamefont {Mohseni}},\
  and\ \bibinfo {author} {\bibfnamefont {P.}~\bibnamefont {Rebentrost}},\
  }\href {https://doi.org/10.1038/nphys3029} {\bibfield  {journal} {\bibinfo
  {journal} {Nat. Phys.}\ }\textbf {\bibinfo {volume} {10}},\ \bibinfo {pages}
  {631} (\bibinfo {year} {2014})}\BibitemShut {NoStop}%
\bibitem [{\citenamefont {Li}\ \emph {et~al.}(2021)\citenamefont {Li},
  \citenamefont {Chai}, \citenamefont {Guo}, \citenamefont {Ji}, \citenamefont
  {Wang}, \citenamefont {Shi}, \citenamefont {Wang}, \citenamefont {Lloyd},\
  and\ \citenamefont {Du}}]{k2005}%
  \BibitemOpen
  \bibfield  {author} {\bibinfo {author} {\bibfnamefont {Z.~K.}\ \bibnamefont
  {Li}}, \bibinfo {author} {\bibfnamefont {Z.~H.}\ \bibnamefont {Chai}},
  \bibinfo {author} {\bibfnamefont {Y.~H.}\ \bibnamefont {Guo}}, \bibinfo
  {author} {\bibfnamefont {W.~T.}\ \bibnamefont {Ji}}, \bibinfo {author}
  {\bibfnamefont {M.~Q.}\ \bibnamefont {Wang}}, \bibinfo {author}
  {\bibfnamefont {F.~Z.}\ \bibnamefont {Shi}}, \bibinfo {author} {\bibfnamefont
  {Y.}~\bibnamefont {Wang}}, \bibinfo {author} {\bibfnamefont {S.}~\bibnamefont
  {Lloyd}},\ and\ \bibinfo {author} {\bibfnamefont {J.~F.}\ \bibnamefont
  {Du}},\ }\href {https://doi.org/10.1126/sciadv.abg2589} {\bibfield  {journal}
  {\bibinfo  {journal} {Sci. Adv.}\ }\textbf {\bibinfo {volume} {7}},\ \bibinfo
  {pages} {eabg2589} (\bibinfo {year} {2021})}\BibitemShut {NoStop}%
\bibitem [{\citenamefont {Tang}(2021)}]{k2006}%
  \BibitemOpen
  \bibfield  {author} {\bibinfo {author} {\bibfnamefont {E.}~\bibnamefont
  {Tang}},\ }\href {https://doi.org/10.1103/PhysRevLett.127.060503} {\bibfield
  {journal} {\bibinfo  {journal} {Phys. Rev. Lett.}\ }\textbf {\bibinfo
  {volume} {127}},\ \bibinfo {pages} {060503} (\bibinfo {year}
  {2021})}\BibitemShut {NoStop}%
\bibitem [{\citenamefont {Gordon}\ \emph {et~al.}(2022)\citenamefont {Gordon},
  \citenamefont {Cerezo}, \citenamefont {Cincio},\ and\ \citenamefont
  {Coles}}]{k2007}%
  \BibitemOpen
  \bibfield  {author} {\bibinfo {author} {\bibfnamefont {M.~H.}\ \bibnamefont
  {Gordon}}, \bibinfo {author} {\bibfnamefont {M.}~\bibnamefont {Cerezo}},
  \bibinfo {author} {\bibfnamefont {L.}~\bibnamefont {Cincio}},\ and\ \bibinfo
  {author} {\bibfnamefont {P.~J.}\ \bibnamefont {Coles}},\ }\href
  {https://doi.org/10.1103/PRXQuantum.3.030334} {\bibfield  {journal} {\bibinfo
   {journal} {PRX Quantum}\ }\textbf {\bibinfo {volume} {3}},\ \bibinfo {pages}
  {030334} (\bibinfo {year} {2022})}\BibitemShut {NoStop}%
\bibitem [{\citenamefont {Rebentrost}\ \emph {et~al.}(2014)\citenamefont
  {Rebentrost}, \citenamefont {Mohseni},\ and\ \citenamefont {Lloyd}}]{k2008}%
  \BibitemOpen
  \bibfield  {author} {\bibinfo {author} {\bibfnamefont {P.}~\bibnamefont
  {Rebentrost}}, \bibinfo {author} {\bibfnamefont {M.}~\bibnamefont
  {Mohseni}},\ and\ \bibinfo {author} {\bibfnamefont {S.}~\bibnamefont
  {Lloyd}},\ }\href {https://doi.org/10.1103/PhysRevLett.113.130503} {\bibfield
   {journal} {\bibinfo  {journal} {Phys. Rev. Lett.}\ }\textbf {\bibinfo
  {volume} {113}},\ \bibinfo {pages} {130503} (\bibinfo {year}
  {2014})}\BibitemShut {NoStop}%
\bibitem [{\citenamefont {Li}\ \emph {et~al.}(2015)\citenamefont {Li},
  \citenamefont {Liu}, \citenamefont {Xu},\ and\ \citenamefont {Du}}]{k2009}%
  \BibitemOpen
  \bibfield  {author} {\bibinfo {author} {\bibfnamefont {Z.~K.}\ \bibnamefont
  {Li}}, \bibinfo {author} {\bibfnamefont {X.~M.}\ \bibnamefont {Liu}},
  \bibinfo {author} {\bibfnamefont {N.~Y.}\ \bibnamefont {Xu}},\ and\ \bibinfo
  {author} {\bibfnamefont {J.~F.}\ \bibnamefont {Du}},\ }\href
  {https://doi.org/10.1103/PhysRevLett.114.140504} {\bibfield  {journal}
  {\bibinfo  {journal} {Phys. Rev. Lett.}\ }\textbf {\bibinfo {volume} {114}},\
  \bibinfo {pages} {140504} (\bibinfo {year} {2015})}\BibitemShut {NoStop}%
\bibitem [{\citenamefont {Gentinetta}\ \emph {et~al.}(2024)\citenamefont
  {Gentinetta}, \citenamefont {Thomsen}, \citenamefont {Sutter},\ and\
  \citenamefont {Woerner}}]{k20010}%
  \BibitemOpen
  \bibfield  {author} {\bibinfo {author} {\bibfnamefont {G.}~\bibnamefont
  {Gentinetta}}, \bibinfo {author} {\bibfnamefont {A.}~\bibnamefont {Thomsen}},
  \bibinfo {author} {\bibfnamefont {D.}~\bibnamefont {Sutter}},\ and\ \bibinfo
  {author} {\bibfnamefont {S.}~\bibnamefont {Woerner}},\ }\href
  {https://doi.org/10.22331/q-2024-01-11-1225} {\bibfield  {journal} {\bibinfo
  {journal} {{Quantum}}\ }\textbf {\bibinfo {volume} {8}},\ \bibinfo {pages}
  {1225} (\bibinfo {year} {2024})}\BibitemShut {NoStop}%
\bibitem [{\citenamefont {Horn}\ and\ \citenamefont {Gottlieb}(2001)}]{k20011}%
  \BibitemOpen
  \bibfield  {author} {\bibinfo {author} {\bibfnamefont {D.}~\bibnamefont
  {Horn}}\ and\ \bibinfo {author} {\bibfnamefont {A.}~\bibnamefont
  {Gottlieb}},\ }\href {https://doi.org/10.1103/PhysRevLett.88.018702}
  {\bibfield  {journal} {\bibinfo  {journal} {Phys. Rev. Lett.}\ }\textbf
  {\bibinfo {volume} {88}},\ \bibinfo {pages} {018702} (\bibinfo {year}
  {2001})}\BibitemShut {NoStop}%
\bibitem [{\citenamefont {Weinstein}\ and\ \citenamefont
  {Horn}(2009)}]{k20012}%
  \BibitemOpen
  \bibfield  {author} {\bibinfo {author} {\bibfnamefont {M.}~\bibnamefont
  {Weinstein}}\ and\ \bibinfo {author} {\bibfnamefont {D.}~\bibnamefont
  {Horn}},\ }\href {https://doi.org/10.1103/PhysRevE.80.066117} {\bibfield
  {journal} {\bibinfo  {journal} {Phys. Rev. E}\ }\textbf {\bibinfo {volume}
  {80}},\ \bibinfo {pages} {066117} (\bibinfo {year} {2009})}\BibitemShut
  {NoStop}%
\bibitem [{\citenamefont {Han}\ \emph {et~al.}(2018)\citenamefont {Han},
  \citenamefont {Wang}, \citenamefont {Fan}, \citenamefont {Wang},\ and\
  \citenamefont {Zhang}}]{k20013}%
  \BibitemOpen
  \bibfield  {author} {\bibinfo {author} {\bibfnamefont {Z.~Y.}\ \bibnamefont
  {Han}}, \bibinfo {author} {\bibfnamefont {J.}~\bibnamefont {Wang}}, \bibinfo
  {author} {\bibfnamefont {H.}~\bibnamefont {Fan}}, \bibinfo {author}
  {\bibfnamefont {L.}~\bibnamefont {Wang}},\ and\ \bibinfo {author}
  {\bibfnamefont {P.}~\bibnamefont {Zhang}},\ }\href
  {https://doi.org/10.1103/PhysRevX.8.031012} {\bibfield  {journal} {\bibinfo
  {journal} {Phys. Rev. X}\ }\textbf {\bibinfo {volume} {8}},\ \bibinfo {pages}
  {031012} (\bibinfo {year} {2018})}\BibitemShut {NoStop}%
\bibitem [{\citenamefont {Or{\'u}s}(2019)}]{kb002001}%
  \BibitemOpen
  \bibfield  {author} {\bibinfo {author} {\bibfnamefont {R.}~\bibnamefont
  {Or{\'u}s}},\ }\href {https://doi.org/10.1038/s42254-019-0086-7} {\bibfield
  {journal} {\bibinfo  {journal} {Nat. Rev. Phys.}\ }\textbf {\bibinfo {volume}
  {1}},\ \bibinfo {pages} {538} (\bibinfo {year} {2019})}\BibitemShut {NoStop}%
\bibitem [{\citenamefont {Lunney}\ \emph {et~al.}(2003)\citenamefont {Lunney},
  \citenamefont {Pearson},\ and\ \citenamefont {Thibault}}]{k3001}%
  \BibitemOpen
  \bibfield  {author} {\bibinfo {author} {\bibfnamefont {D.}~\bibnamefont
  {Lunney}}, \bibinfo {author} {\bibfnamefont {J.~M.}\ \bibnamefont
  {Pearson}},\ and\ \bibinfo {author} {\bibfnamefont {C.}~\bibnamefont
  {Thibault}},\ }\href {https://doi.org/10.1103/RevModPhys.75.1021} {\bibfield
  {journal} {\bibinfo  {journal} {Rev. Mod. Phys.}\ }\textbf {\bibinfo {volume}
  {75}},\ \bibinfo {pages} {1021} (\bibinfo {year} {2003})}\BibitemShut
  {NoStop}%
\bibitem [{\citenamefont {Erler}\ \emph {et~al.}(2012)\citenamefont {Erler},
  \citenamefont {Birge}, \citenamefont {Kortelainen}, \citenamefont
  {Nazarewicz}, \citenamefont {Olsen}, \citenamefont {Perhac},\ and\
  \citenamefont {Stoitsov}}]{k3002}%
  \BibitemOpen
  \bibfield  {author} {\bibinfo {author} {\bibfnamefont {J.}~\bibnamefont
  {Erler}}, \bibinfo {author} {\bibfnamefont {N.}~\bibnamefont {Birge}},
  \bibinfo {author} {\bibfnamefont {M.}~\bibnamefont {Kortelainen}}, \bibinfo
  {author} {\bibfnamefont {W.}~\bibnamefont {Nazarewicz}}, \bibinfo {author}
  {\bibfnamefont {E.}~\bibnamefont {Olsen}}, \bibinfo {author} {\bibfnamefont
  {A.~M.}\ \bibnamefont {Perhac}},\ and\ \bibinfo {author} {\bibfnamefont
  {M.}~\bibnamefont {Stoitsov}},\ }\href {https://doi.org/10.1038/nature11188}
  {\bibfield  {journal} {\bibinfo  {journal} {Nature}\ }\textbf {\bibinfo
  {volume} {486}},\ \bibinfo {pages} {509} (\bibinfo {year}
  {2012})}\BibitemShut {NoStop}%
\bibitem [{\citenamefont {Yu}\ \emph {et~al.}(2024)\citenamefont {Yu},
  \citenamefont {Xing}, \citenamefont {Zhang}, \citenamefont {Wang},
  \citenamefont {Zhou}, \citenamefont {Li}, \citenamefont {Li}, \citenamefont
  {Yuan}, \citenamefont {Niu}, \citenamefont {Huang}, \citenamefont {Geng},
  \citenamefont {Guo}, \citenamefont {Chen}, \citenamefont {Pei}, \citenamefont
  {Xu}, \citenamefont {Litvinov}, \citenamefont {Blaum}, \citenamefont
  {de~Angelis}, \citenamefont {Tanihata}, \citenamefont {Yamaguchi},
  \citenamefont {Zhou}, \citenamefont {Xu}, \citenamefont {Chen}, \citenamefont
  {Chen}, \citenamefont {Deng}, \citenamefont {Fu}, \citenamefont {Ge},
  \citenamefont {Huang}, \citenamefont {Jiao}, \citenamefont {Luo},
  \citenamefont {Li}, \citenamefont {Liao}, \citenamefont {Shi}, \citenamefont
  {Si}, \citenamefont {Sun}, \citenamefont {Shuai}, \citenamefont {Tu},
  \citenamefont {Wang}, \citenamefont {Xu}, \citenamefont {Yan}, \citenamefont
  {Yuan},\ and\ \citenamefont {Zhang}}]{k3003}%
  \BibitemOpen
  \bibfield  {author} {\bibinfo {author} {\bibfnamefont {Y.}~\bibnamefont
  {Yu}}, \bibinfo {author} {\bibfnamefont {Y.~M.}\ \bibnamefont {Xing}},
  \bibinfo {author} {\bibfnamefont {Y.~H.}\ \bibnamefont {Zhang}}, \bibinfo
  {author} {\bibfnamefont {M.}~\bibnamefont {Wang}}, \bibinfo {author}
  {\bibfnamefont {X.~H.}\ \bibnamefont {Zhou}}, \bibinfo {author}
  {\bibfnamefont {J.~G.}\ \bibnamefont {Li}}, \bibinfo {author} {\bibfnamefont
  {H.~H.}\ \bibnamefont {Li}}, \bibinfo {author} {\bibfnamefont
  {Q.}~\bibnamefont {Yuan}}, \bibinfo {author} {\bibfnamefont {Y.~F.}\
  \bibnamefont {Niu}}, \bibinfo {author} {\bibfnamefont {Y.~N.}\ \bibnamefont
  {Huang}}, \emph {et~al.},\ }\href
  {https://doi.org/10.1103/PhysRevLett.133.222501} {\bibfield  {journal}
  {\bibinfo  {journal} {Phys. Rev. Lett.}\ }\textbf {\bibinfo {volume} {133}},\
  \bibinfo {pages} {222501} (\bibinfo {year} {2024})}\BibitemShut {NoStop}%
\bibitem [{\citenamefont {Otsuka}\ \emph {et~al.}(2020)\citenamefont {Otsuka},
  \citenamefont {Gade}, \citenamefont {Sorlin}, \citenamefont {Suzuki},\ and\
  \citenamefont {Utsuno}}]{k3004}%
  \BibitemOpen
  \bibfield  {author} {\bibinfo {author} {\bibfnamefont {T.}~\bibnamefont
  {Otsuka}}, \bibinfo {author} {\bibfnamefont {A.}~\bibnamefont {Gade}},
  \bibinfo {author} {\bibfnamefont {O.}~\bibnamefont {Sorlin}}, \bibinfo
  {author} {\bibfnamefont {T.}~\bibnamefont {Suzuki}},\ and\ \bibinfo {author}
  {\bibfnamefont {Y.}~\bibnamefont {Utsuno}},\ }\href
  {https://doi.org/10.1103/RevModPhys.92.015002} {\bibfield  {journal}
  {\bibinfo  {journal} {Rev. Mod. Phys.}\ }\textbf {\bibinfo {volume} {92}},\
  \bibinfo {pages} {015002} (\bibinfo {year} {2020})}\BibitemShut {NoStop}%
\bibitem [{\citenamefont {Gelberg}\ \emph {et~al.}(2009)\citenamefont
  {Gelberg}, \citenamefont {Sakurai}, \citenamefont {Kirson},\ and\
  \citenamefont {Heinze}}]{3.e1}%
  \BibitemOpen
  \bibfield  {author} {\bibinfo {author} {\bibfnamefont {A.}~\bibnamefont
  {Gelberg}}, \bibinfo {author} {\bibfnamefont {H.}~\bibnamefont {Sakurai}},
  \bibinfo {author} {\bibfnamefont {M.~W.}\ \bibnamefont {Kirson}},\ and\
  \bibinfo {author} {\bibfnamefont {S.}~\bibnamefont {Heinze}},\ }\href
  {https://doi.org/10.1103/PhysRevC.80.024307} {\bibfield  {journal} {\bibinfo
  {journal} {Phys. Rev. C}\ }\textbf {\bibinfo {volume} {80}},\ \bibinfo
  {pages} {024307} (\bibinfo {year} {2009})}\BibitemShut {NoStop}%
\bibitem [{\citenamefont {Cowan}\ \emph {et~al.}(2021)\citenamefont {Cowan},
  \citenamefont {Sneden}, \citenamefont {Lawler}, \citenamefont {Aprahamian},
  \citenamefont {Wiescher}, \citenamefont {Langanke}, \citenamefont
  {Mart\'{\i}nez-Pinedo},\ and\ \citenamefont {Thielemann}}]{k3005}%
  \BibitemOpen
  \bibfield  {author} {\bibinfo {author} {\bibfnamefont {J.~J.}\ \bibnamefont
  {Cowan}}, \bibinfo {author} {\bibfnamefont {C.}~\bibnamefont {Sneden}},
  \bibinfo {author} {\bibfnamefont {J.~E.}\ \bibnamefont {Lawler}}, \bibinfo
  {author} {\bibfnamefont {A.}~\bibnamefont {Aprahamian}}, \bibinfo {author}
  {\bibfnamefont {M.}~\bibnamefont {Wiescher}}, \bibinfo {author}
  {\bibfnamefont {K.}~\bibnamefont {Langanke}}, \bibinfo {author}
  {\bibfnamefont {G.}~\bibnamefont {Mart\'{\i}nez-Pinedo}},\ and\ \bibinfo
  {author} {\bibfnamefont {F.~K.}\ \bibnamefont {Thielemann}},\ }\href
  {https://doi.org/10.1103/RevModPhys.93.015002} {\bibfield  {journal}
  {\bibinfo  {journal} {Rev. Mod. Phys.}\ }\textbf {\bibinfo {volume} {93}},\
  \bibinfo {pages} {015002} (\bibinfo {year} {2021})}\BibitemShut {NoStop}%
\bibitem [{\citenamefont {Williams}\ \emph {et~al.}(2025)\citenamefont
  {Williams}, \citenamefont {Angus}, \citenamefont {Laird}, \citenamefont
  {Davids}, \citenamefont {Diget}, \citenamefont {Fernandez}, \citenamefont
  {Williams}, \citenamefont {Andreyev}, \citenamefont {Asch}, \citenamefont
  {Avaa}, \citenamefont {Bartram}, \citenamefont {Chakraborty}, \citenamefont
  {Dillmann}, \citenamefont {Directo}, \citenamefont {Doherty}, \citenamefont
  {Geerlof}, \citenamefont {Griffin}, \citenamefont {Grimes}, \citenamefont
  {Hackman}, \citenamefont {Henderson}, \citenamefont {Hudson}, \citenamefont
  {Hufschmidt}, \citenamefont {Jeong}, \citenamefont {Jim\'enez~de Haro},
  \citenamefont {Karayonchev}, \citenamefont {Katrusiak}, \citenamefont
  {Lennarz}, \citenamefont {Lotay}, \citenamefont {Marlow}, \citenamefont
  {Martin}, \citenamefont {Moll\'o}, \citenamefont {Montes}, \citenamefont
  {Murias}, \citenamefont {O'Neill}, \citenamefont {Pak}, \citenamefont
  {Paxman}, \citenamefont {Pedro-Botet}, \citenamefont {Psaltis}, \citenamefont
  {Raleigh-Smith}, \citenamefont {Rhodes}, \citenamefont {Rojo}, \citenamefont
  {Satrazani}, \citenamefont {Sauvage}, \citenamefont {Shenton}, \citenamefont
  {Svensson}, \citenamefont {Tam}, \citenamefont {Wagner},\ and\ \citenamefont
  {Yates}}]{k3006}%
  \BibitemOpen
  \bibfield  {author} {\bibinfo {author} {\bibfnamefont {M.}~\bibnamefont
  {Williams}}, \bibinfo {author} {\bibfnamefont {C.}~\bibnamefont {Angus}},
  \bibinfo {author} {\bibfnamefont {A.~M.}\ \bibnamefont {Laird}}, \bibinfo
  {author} {\bibfnamefont {B.}~\bibnamefont {Davids}}, \bibinfo {author}
  {\bibfnamefont {C.~A.}\ \bibnamefont {Diget}}, \bibinfo {author}
  {\bibfnamefont {A.}~\bibnamefont {Fernandez}}, \bibinfo {author}
  {\bibfnamefont {E.~J.}\ \bibnamefont {Williams}}, \bibinfo {author}
  {\bibfnamefont {A.~N.}\ \bibnamefont {Andreyev}}, \bibinfo {author}
  {\bibfnamefont {H.}~\bibnamefont {Asch}}, \bibinfo {author} {\bibfnamefont
  {A.~A.}\ \bibnamefont {Avaa}}, \emph {et~al.},\ }\href
  {https://doi.org/10.1103/PhysRevLett.134.112701} {\bibfield  {journal}
  {\bibinfo  {journal} {Phys. Rev. Lett.}\ }\textbf {\bibinfo {volume} {134}},\
  \bibinfo {pages} {112701} (\bibinfo {year} {2025})}\BibitemShut {NoStop}%
\bibitem [{\citenamefont {Kl\"upfel}\ \emph {et~al.}(2009)\citenamefont
  {Kl\"upfel}, \citenamefont {Reinhard}, \citenamefont {B\"urvenich},\ and\
  \citenamefont {Maruhn}}]{k3007}%
  \BibitemOpen
  \bibfield  {author} {\bibinfo {author} {\bibfnamefont {P.}~\bibnamefont
  {Kl\"upfel}}, \bibinfo {author} {\bibfnamefont {P.-G.}\ \bibnamefont
  {Reinhard}}, \bibinfo {author} {\bibfnamefont {T.~J.}\ \bibnamefont
  {B\"urvenich}},\ and\ \bibinfo {author} {\bibfnamefont {J.~A.}\ \bibnamefont
  {Maruhn}},\ }\href {https://doi.org/10.1103/PhysRevC.79.034310} {\bibfield
  {journal} {\bibinfo  {journal} {Phys. Rev. C}\ }\textbf {\bibinfo {volume}
  {79}},\ \bibinfo {pages} {034310} (\bibinfo {year} {2009})}\BibitemShut
  {NoStop}%
\bibitem [{\citenamefont {Kortelainen}\ \emph {et~al.}(2010)\citenamefont
  {Kortelainen}, \citenamefont {Lesinski}, \citenamefont {Mor\'e},
  \citenamefont {Nazarewicz}, \citenamefont {Sarich}, \citenamefont {Schunck},
  \citenamefont {Stoitsov},\ and\ \citenamefont {Wild}}]{k3008}%
  \BibitemOpen
  \bibfield  {author} {\bibinfo {author} {\bibfnamefont {M.}~\bibnamefont
  {Kortelainen}}, \bibinfo {author} {\bibfnamefont {T.}~\bibnamefont
  {Lesinski}}, \bibinfo {author} {\bibfnamefont {J.}~\bibnamefont {Mor\'e}},
  \bibinfo {author} {\bibfnamefont {W.}~\bibnamefont {Nazarewicz}}, \bibinfo
  {author} {\bibfnamefont {J.}~\bibnamefont {Sarich}}, \bibinfo {author}
  {\bibfnamefont {N.}~\bibnamefont {Schunck}}, \bibinfo {author} {\bibfnamefont
  {M.~V.}\ \bibnamefont {Stoitsov}},\ and\ \bibinfo {author} {\bibfnamefont
  {S.}~\bibnamefont {Wild}},\ }\href
  {https://doi.org/10.1103/PhysRevC.82.024313} {\bibfield  {journal} {\bibinfo
  {journal} {Phys. Rev. C}\ }\textbf {\bibinfo {volume} {82}},\ \bibinfo
  {pages} {024313} (\bibinfo {year} {2010})}\BibitemShut {NoStop}%
\bibitem [{\citenamefont {Agbemava}\ \emph
  {et~al.}(2014{\natexlab{a}})\citenamefont {Agbemava}, \citenamefont
  {Afanasjev}, \citenamefont {Ray},\ and\ \citenamefont {Ring}}]{k3009}%
  \BibitemOpen
  \bibfield  {author} {\bibinfo {author} {\bibfnamefont {S.~E.}\ \bibnamefont
  {Agbemava}}, \bibinfo {author} {\bibfnamefont {A.~V.}\ \bibnamefont
  {Afanasjev}}, \bibinfo {author} {\bibfnamefont {D.}~\bibnamefont {Ray}},\
  and\ \bibinfo {author} {\bibfnamefont {P.}~\bibnamefont {Ring}},\ }\href
  {https://doi.org/10.1103/PhysRevC.89.054320} {\bibfield  {journal} {\bibinfo
  {journal} {Phys. Rev. C}\ }\textbf {\bibinfo {volume} {89}},\ \bibinfo
  {pages} {054320} (\bibinfo {year} {2014}{\natexlab{a}})}\BibitemShut
  {NoStop}%
\bibitem [{\citenamefont {Afanasjev}\ and\ \citenamefont
  {Agbemava}(2016)}]{k30010}%
  \BibitemOpen
  \bibfield  {author} {\bibinfo {author} {\bibfnamefont {A.~V.}\ \bibnamefont
  {Afanasjev}}\ and\ \bibinfo {author} {\bibfnamefont {S.~E.}\ \bibnamefont
  {Agbemava}},\ }\href {https://doi.org/10.1103/PhysRevC.93.054310} {\bibfield
  {journal} {\bibinfo  {journal} {Phys. Rev. C}\ }\textbf {\bibinfo {volume}
  {93}},\ \bibinfo {pages} {054310} (\bibinfo {year} {2016})}\BibitemShut
  {NoStop}%
\bibitem [{\citenamefont {Wolfgruber}\ \emph {et~al.}(2024)\citenamefont
  {Wolfgruber}, \citenamefont {Kn\"oll},\ and\ \citenamefont {Roth}}]{k30011}%
  \BibitemOpen
  \bibfield  {author} {\bibinfo {author} {\bibfnamefont {T.}~\bibnamefont
  {Wolfgruber}}, \bibinfo {author} {\bibfnamefont {M.}~\bibnamefont
  {Kn\"oll}},\ and\ \bibinfo {author} {\bibfnamefont {R.}~\bibnamefont
  {Roth}},\ }\href {https://doi.org/10.1103/PhysRevC.110.014327} {\bibfield
  {journal} {\bibinfo  {journal} {Phys. Rev. C}\ }\textbf {\bibinfo {volume}
  {110}},\ \bibinfo {pages} {014327} (\bibinfo {year} {2024})}\BibitemShut
  {NoStop}%
\bibitem [{\citenamefont {Guo}\ \emph {et~al.}(2024)\citenamefont {Guo},
  \citenamefont {Cao}, \citenamefont {Chen}, \citenamefont {Chen},
  \citenamefont {Cheoun}, \citenamefont {Choi}, \citenamefont {Lam},
  \citenamefont {Deng}, \citenamefont {Dong}, \citenamefont {Du}, \citenamefont
  {Du}, \citenamefont {Duan}, \citenamefont {Fan}, \citenamefont {Gao},
  \citenamefont {Geng}, \citenamefont {Ha}, \citenamefont {He}, \citenamefont
  {Hu}, \citenamefont {Huang}, \citenamefont {Huang}, \citenamefont {Huang},
  \citenamefont {Huang}, \citenamefont {Hyung}, \citenamefont {Chan},
  \citenamefont {Jiang}, \citenamefont {Kim}, \citenamefont {Kim},
  \citenamefont {Lee}, \citenamefont {Lee}, \citenamefont {Li}, \citenamefont
  {Li}, \citenamefont {Li}, \citenamefont {Li}, \citenamefont {Lian},
  \citenamefont {Liang}, \citenamefont {Liu}, \citenamefont {Lu}, \citenamefont
  {Liu}, \citenamefont {Meng}, \citenamefont {Meng}, \citenamefont {Mun},
  \citenamefont {Niu}, \citenamefont {Niu}, \citenamefont {Pan}, \citenamefont
  {Peng}, \citenamefont {Qu}, \citenamefont {Papakonstantinou}, \citenamefont
  {Shang}, \citenamefont {Shang}, \citenamefont {Shen}, \citenamefont {Shen},
  \citenamefont {Sun}, \citenamefont {Sun}, \citenamefont {Wang}, \citenamefont
  {Wang}, \citenamefont {Wang}, \citenamefont {Wang}, \citenamefont {Wu},
  \citenamefont {Wu}, \citenamefont {Wu}, \citenamefont {Xia}, \citenamefont
  {Xie}, \citenamefont {Yao}, \citenamefont {Ip}, \citenamefont {Yiu},
  \citenamefont {Yu}, \citenamefont {Yu}, \citenamefont {Zhang}, \citenamefont
  {Zhang}, \citenamefont {Zhang}, \citenamefont {Zhang}, \citenamefont {Zhang},
  \citenamefont {Zhang}, \citenamefont {Zhang}, \citenamefont {Zhang},
  \citenamefont {Zhang}, \citenamefont {Zhao}, \citenamefont {Zhao},
  \citenamefont {Zheng}, \citenamefont {Zhou}, \citenamefont {Zhou},\ and\
  \citenamefont {Zou}}]{k30012}%
  \BibitemOpen
  \bibfield  {author} {\bibinfo {author} {\bibfnamefont {P.}~\bibnamefont
  {Guo}}, \bibinfo {author} {\bibfnamefont {X.~J.}\ \bibnamefont {Cao}},
  \bibinfo {author} {\bibfnamefont {K.~M.}\ \bibnamefont {Chen}}, \bibinfo
  {author} {\bibfnamefont {Z.~H.}\ \bibnamefont {Chen}}, \bibinfo {author}
  {\bibfnamefont {M.~K.}\ \bibnamefont {Cheoun}}, \bibinfo {author}
  {\bibfnamefont {Y.~B.}\ \bibnamefont {Choi}}, \bibinfo {author}
  {\bibfnamefont {P.~C.}\ \bibnamefont {Lam}}, \bibinfo {author} {\bibfnamefont
  {W.}~\bibnamefont {Deng}}, \bibinfo {author} {\bibfnamefont {J.}~\bibnamefont
  {Dong}}, \bibinfo {author} {\bibfnamefont {P.}~\bibnamefont {Du}}, \emph
  {et~al.},\ }\href {https://doi.org/https://doi.org/10.1016/j.adt.2024.101661}
  {\bibfield  {journal} {\bibinfo  {journal} {At. Data Nucl. Data Tables}\
  }\textbf {\bibinfo {volume} {158}},\ \bibinfo {pages} {101661} (\bibinfo
  {year} {2024})}\BibitemShut {NoStop}%
\bibitem [{\citenamefont {Becker}\ \emph {et~al.}(2017)\citenamefont {Becker},
  \citenamefont {Davesne}, \citenamefont {Meyer}, \citenamefont {Navarro},\
  and\ \citenamefont {Pastore}}]{3.e2}%
  \BibitemOpen
  \bibfield  {author} {\bibinfo {author} {\bibfnamefont {P.}~\bibnamefont
  {Becker}}, \bibinfo {author} {\bibfnamefont {D.}~\bibnamefont {Davesne}},
  \bibinfo {author} {\bibfnamefont {J.}~\bibnamefont {Meyer}}, \bibinfo
  {author} {\bibfnamefont {J.}~\bibnamefont {Navarro}},\ and\ \bibinfo {author}
  {\bibfnamefont {A.}~\bibnamefont {Pastore}},\ }\href
  {https://doi.org/10.1103/PhysRevC.96.044330} {\bibfield  {journal} {\bibinfo
  {journal} {Phys. Rev. C}\ }\textbf {\bibinfo {volume} {96}},\ \bibinfo
  {pages} {044330} (\bibinfo {year} {2017})}\BibitemShut {NoStop}%
\bibitem [{\citenamefont {Geng}\ \emph
  {et~al.}(2025{\natexlab{a}})\citenamefont {Geng}, \citenamefont {Wang},
  \citenamefont {Zhao}, \citenamefont {Niu}, \citenamefont {Liang},\ and\
  \citenamefont {Long}}]{k300b16}%
  \BibitemOpen
  \bibfield  {author} {\bibinfo {author} {\bibfnamefont {J.}~\bibnamefont
  {Geng}}, \bibinfo {author} {\bibfnamefont {Z.~H.}\ \bibnamefont {Wang}},
  \bibinfo {author} {\bibfnamefont {P.~W.}\ \bibnamefont {Zhao}}, \bibinfo
  {author} {\bibfnamefont {Y.~F.}\ \bibnamefont {Niu}}, \bibinfo {author}
  {\bibfnamefont {H.}~\bibnamefont {Liang}},\ and\ \bibinfo {author}
  {\bibfnamefont {W.~H.}\ \bibnamefont {Long}},\ }\href
  {https://doi.org/10.1103/PhysRevC.111.024305} {\bibfield  {journal} {\bibinfo
   {journal} {Phys. Rev. C}\ }\textbf {\bibinfo {volume} {111}},\ \bibinfo
  {pages} {024305} (\bibinfo {year} {2025}{\natexlab{a}})}\BibitemShut
  {NoStop}%
\bibitem [{\citenamefont {Garvey}\ \emph {et~al.}(1969)\citenamefont {Garvey},
  \citenamefont {Gerace}, \citenamefont {Jaffe}, \citenamefont {Talmi},\ and\
  \citenamefont {Kelson}}]{k300b17}%
  \BibitemOpen
  \bibfield  {author} {\bibinfo {author} {\bibfnamefont {G.~T.}\ \bibnamefont
  {Garvey}}, \bibinfo {author} {\bibfnamefont {W.~J.}\ \bibnamefont {Gerace}},
  \bibinfo {author} {\bibfnamefont {R.~L.}\ \bibnamefont {Jaffe}}, \bibinfo
  {author} {\bibfnamefont {I.}~\bibnamefont {Talmi}},\ and\ \bibinfo {author}
  {\bibfnamefont {I.}~\bibnamefont {Kelson}},\ }\href
  {https://doi.org/10.1103/RevModPhys.41.S1} {\bibfield  {journal} {\bibinfo
  {journal} {Rev. Mod. Phys.}\ }\textbf {\bibinfo {volume} {41}},\ \bibinfo
  {pages} {S1} (\bibinfo {year} {1969})}\BibitemShut {NoStop}%
\bibitem [{\citenamefont {Bao}\ \emph {et~al.}(2013)\citenamefont {Bao},
  \citenamefont {He}, \citenamefont {Lu}, \citenamefont {Zhao},\ and\
  \citenamefont {Arima}}]{k30013}%
  \BibitemOpen
  \bibfield  {author} {\bibinfo {author} {\bibfnamefont {M.}~\bibnamefont
  {Bao}}, \bibinfo {author} {\bibfnamefont {Z.}~\bibnamefont {He}}, \bibinfo
  {author} {\bibfnamefont {Y.}~\bibnamefont {Lu}}, \bibinfo {author}
  {\bibfnamefont {Y.~M.}\ \bibnamefont {Zhao}},\ and\ \bibinfo {author}
  {\bibfnamefont {A.}~\bibnamefont {Arima}},\ }\href
  {https://doi.org/10.1103/PhysRevC.88.064325} {\bibfield  {journal} {\bibinfo
  {journal} {Phys. Rev. C}\ }\textbf {\bibinfo {volume} {88}},\ \bibinfo
  {pages} {064325} (\bibinfo {year} {2013})}\BibitemShut {NoStop}%
\bibitem [{\citenamefont {Agbemava}\ \emph
  {et~al.}(2014{\natexlab{b}})\citenamefont {Agbemava}, \citenamefont
  {Afanasjev}, \citenamefont {Ray},\ and\ \citenamefont {Ring}}]{3.e3}%
  \BibitemOpen
  \bibfield  {author} {\bibinfo {author} {\bibfnamefont {S.~E.}\ \bibnamefont
  {Agbemava}}, \bibinfo {author} {\bibfnamefont {A.~V.}\ \bibnamefont
  {Afanasjev}}, \bibinfo {author} {\bibfnamefont {D.}~\bibnamefont {Ray}},\
  and\ \bibinfo {author} {\bibfnamefont {P.}~\bibnamefont {Ring}},\ }\href
  {https://doi.org/10.1103/PhysRevC.89.054320} {\bibfield  {journal} {\bibinfo
  {journal} {Phys. Rev. C}\ }\textbf {\bibinfo {volume} {89}},\ \bibinfo
  {pages} {054320} (\bibinfo {year} {2014}{\natexlab{b}})}\BibitemShut
  {NoStop}%
\bibitem [{\citenamefont {Boehnlein}\ \emph {et~al.}(2022)\citenamefont
  {Boehnlein}, \citenamefont {Diefenthaler}, \citenamefont {Sato},
  \citenamefont {Schram}, \citenamefont {Ziegler}, \citenamefont {Fanelli},
  \citenamefont {Hjorth~Jensen}, \citenamefont {Horn}, \citenamefont {Kuchera},
  \citenamefont {Lee}, \citenamefont {Nazarewicz}, \citenamefont {Ostroumov},
  \citenamefont {Orginos}, \citenamefont {Poon}, \citenamefont {Wang},
  \citenamefont {Scheinker}, \citenamefont {Smith},\ and\ \citenamefont
  {Pang}}]{k1003}%
  \BibitemOpen
  \bibfield  {author} {\bibinfo {author} {\bibfnamefont {A.}~\bibnamefont
  {Boehnlein}}, \bibinfo {author} {\bibfnamefont {M.}~\bibnamefont
  {Diefenthaler}}, \bibinfo {author} {\bibfnamefont {N.}~\bibnamefont {Sato}},
  \bibinfo {author} {\bibfnamefont {M.}~\bibnamefont {Schram}}, \bibinfo
  {author} {\bibfnamefont {V.}~\bibnamefont {Ziegler}}, \bibinfo {author}
  {\bibfnamefont {C.}~\bibnamefont {Fanelli}}, \bibinfo {author} {\bibfnamefont
  {M.}~\bibnamefont {Hjorth~Jensen}}, \bibinfo {author} {\bibfnamefont
  {T.}~\bibnamefont {Horn}}, \bibinfo {author} {\bibfnamefont {M.~P.}\
  \bibnamefont {Kuchera}}, \bibinfo {author} {\bibfnamefont {D.}~\bibnamefont
  {Lee}}, \emph {et~al.},\ }\href
  {https://doi.org/10.1103/RevModPhys.94.031003} {\bibfield  {journal}
  {\bibinfo  {journal} {Rev. Mod. Phys.}\ }\textbf {\bibinfo {volume} {94}},\
  \bibinfo {pages} {031003} (\bibinfo {year} {2022})}\BibitemShut {NoStop}%
\bibitem [{\citenamefont {He}\ \emph {et~al.}(2023)\citenamefont {He},
  \citenamefont {Li}, \citenamefont {Ma}, \citenamefont {Niu}, \citenamefont
  {Pei},\ and\ \citenamefont {Zhang}}]{k1004}%
  \BibitemOpen
  \bibfield  {author} {\bibinfo {author} {\bibfnamefont {W.~B.}\ \bibnamefont
  {He}}, \bibinfo {author} {\bibfnamefont {Q.~F.}\ \bibnamefont {Li}}, \bibinfo
  {author} {\bibfnamefont {Y.~G.}\ \bibnamefont {Ma}}, \bibinfo {author}
  {\bibfnamefont {Z.~M.}\ \bibnamefont {Niu}}, \bibinfo {author} {\bibfnamefont
  {J.~C.}\ \bibnamefont {Pei}},\ and\ \bibinfo {author} {\bibfnamefont {Y.~X.}\
  \bibnamefont {Zhang}},\ }\href {https://doi.org/10.1007/s11433-023-2116-0}
  {\bibfield  {journal} {\bibinfo  {journal} {Sci. China Phys. Mech. Astron.}\
  }\textbf {\bibinfo {volume} {66}},\ \bibinfo {pages} {282001} (\bibinfo
  {year} {2023})}\BibitemShut {NoStop}%
\bibitem [{\citenamefont {Gernoth}\ \emph {et~al.}(1993)\citenamefont
  {Gernoth}, \citenamefont {Clark}, \citenamefont {Prater},\ and\ \citenamefont
  {Bohr}}]{3.d1}%
  \BibitemOpen
  \bibfield  {author} {\bibinfo {author} {\bibfnamefont {K.}~\bibnamefont
  {Gernoth}}, \bibinfo {author} {\bibfnamefont {J.}~\bibnamefont {Clark}},
  \bibinfo {author} {\bibfnamefont {J.}~\bibnamefont {Prater}},\ and\ \bibinfo
  {author} {\bibfnamefont {H.}~\bibnamefont {Bohr}},\ }\href
  {https://doi.org/https://doi.org/10.1016/0370-2693(93)90738-4} {\bibfield
  {journal} {\bibinfo  {journal} {Phys. Lett. B}\ }\textbf {\bibinfo {volume}
  {300}},\ \bibinfo {pages} {1} (\bibinfo {year} {1993})}\BibitemShut {NoStop}%
\bibitem [{\citenamefont {Utama}\ \emph {et~al.}(2016)\citenamefont {Utama},
  \citenamefont {Piekarewicz},\ and\ \citenamefont {Prosper}}]{3.d2}%
  \BibitemOpen
  \bibfield  {author} {\bibinfo {author} {\bibfnamefont {R.}~\bibnamefont
  {Utama}}, \bibinfo {author} {\bibfnamefont {J.}~\bibnamefont {Piekarewicz}},\
  and\ \bibinfo {author} {\bibfnamefont {H.~B.}\ \bibnamefont {Prosper}},\
  }\href {https://doi.org/10.1103/PhysRevC.93.014311} {\bibfield  {journal}
  {\bibinfo  {journal} {Phys. Rev. C}\ }\textbf {\bibinfo {volume} {93}},\
  \bibinfo {pages} {014311} (\bibinfo {year} {2016})}\BibitemShut {NoStop}%
\bibitem [{\citenamefont {Niu}\ \emph {et~al.}(2019{\natexlab{a}})\citenamefont
  {Niu}, \citenamefont {Fang},\ and\ \citenamefont {Niu}}]{k300b2}%
  \BibitemOpen
  \bibfield  {author} {\bibinfo {author} {\bibfnamefont {Z.~M.}\ \bibnamefont
  {Niu}}, \bibinfo {author} {\bibfnamefont {J.~Y.}\ \bibnamefont {Fang}},\ and\
  \bibinfo {author} {\bibfnamefont {Y.~F.}\ \bibnamefont {Niu}},\ }\href
  {https://doi.org/10.1103/PhysRevC.100.054311} {\bibfield  {journal} {\bibinfo
   {journal} {Phys. Rev. C}\ }\textbf {\bibinfo {volume} {100}},\ \bibinfo
  {pages} {054311} (\bibinfo {year} {2019}{\natexlab{a}})}\BibitemShut
  {NoStop}%
\bibitem [{\citenamefont {Lovell}\ \emph {et~al.}(2022)\citenamefont {Lovell},
  \citenamefont {Mohan}, \citenamefont {Sprouse},\ and\ \citenamefont
  {Mumpower}}]{k300b3}%
  \BibitemOpen
  \bibfield  {author} {\bibinfo {author} {\bibfnamefont {A.~E.}\ \bibnamefont
  {Lovell}}, \bibinfo {author} {\bibfnamefont {A.~T.}\ \bibnamefont {Mohan}},
  \bibinfo {author} {\bibfnamefont {T.~M.}\ \bibnamefont {Sprouse}},\ and\
  \bibinfo {author} {\bibfnamefont {M.~R.}\ \bibnamefont {Mumpower}},\ }\href
  {https://doi.org/10.1103/PhysRevC.106.014305} {\bibfield  {journal} {\bibinfo
   {journal} {Phys. Rev. C}\ }\textbf {\bibinfo {volume} {106}},\ \bibinfo
  {pages} {014305} (\bibinfo {year} {2022})}\BibitemShut {NoStop}%
\bibitem [{\citenamefont {Sharma}\ \emph {et~al.}(2022)\citenamefont {Sharma},
  \citenamefont {Gandhi},\ and\ \citenamefont {Kumar}}]{k300b4}%
  \BibitemOpen
  \bibfield  {author} {\bibinfo {author} {\bibfnamefont {A.}~\bibnamefont
  {Sharma}}, \bibinfo {author} {\bibfnamefont {A.}~\bibnamefont {Gandhi}},\
  and\ \bibinfo {author} {\bibfnamefont {A.}~\bibnamefont {Kumar}},\ }\href
  {https://doi.org/10.1103/PhysRevC.105.L031306} {\bibfield  {journal}
  {\bibinfo  {journal} {Phys. Rev. C}\ }\textbf {\bibinfo {volume} {105}},\
  \bibinfo {pages} {L031306} (\bibinfo {year} {2022})}\BibitemShut {NoStop}%
\bibitem [{\citenamefont {Lu}\ \emph {et~al.}(2025)\citenamefont {Lu},
  \citenamefont {Shang}, \citenamefont {Du}, \citenamefont {Li}, \citenamefont
  {Liang},\ and\ \citenamefont {Niu}}]{k300b5}%
  \BibitemOpen
  \bibfield  {author} {\bibinfo {author} {\bibfnamefont {Y.~H.}\ \bibnamefont
  {Lu}}, \bibinfo {author} {\bibfnamefont {T.~S.}\ \bibnamefont {Shang}},
  \bibinfo {author} {\bibfnamefont {P.~X.}\ \bibnamefont {Du}}, \bibinfo
  {author} {\bibfnamefont {J.}~\bibnamefont {Li}}, \bibinfo {author}
  {\bibfnamefont {H.~Z.}\ \bibnamefont {Liang}},\ and\ \bibinfo {author}
  {\bibfnamefont {Z.~M.}\ \bibnamefont {Niu}},\ }\href
  {https://doi.org/10.1103/PhysRevC.111.014325} {\bibfield  {journal} {\bibinfo
   {journal} {Phys. Rev. C}\ }\textbf {\bibinfo {volume} {111}},\ \bibinfo
  {pages} {014325} (\bibinfo {year} {2025})}\BibitemShut {NoStop}%
\bibitem [{\citenamefont {Gao}\ \emph {et~al.}(2021)\citenamefont {Gao},
  \citenamefont {Wang}, \citenamefont {Lü}, \citenamefont {Li}, \citenamefont
  {Shen},\ and\ \citenamefont {Liu}}]{k300b8}%
  \BibitemOpen
  \bibfield  {author} {\bibinfo {author} {\bibfnamefont {Z.~P.}\ \bibnamefont
  {Gao}}, \bibinfo {author} {\bibfnamefont {Y.~J.}\ \bibnamefont {Wang}},
  \bibinfo {author} {\bibfnamefont {H.~L.}\ \bibnamefont {Lü}}, \bibinfo
  {author} {\bibfnamefont {Q.~F.}\ \bibnamefont {Li}}, \bibinfo {author}
  {\bibfnamefont {C.~W.}\ \bibnamefont {Shen}},\ and\ \bibinfo {author}
  {\bibfnamefont {L.}~\bibnamefont {Liu}},\ }\href
  {https://doi.org/10.1007/s41365-021-00956-1} {\bibfield  {journal} {\bibinfo
  {journal} {Nucl. Sci. Tech.}\ }\textbf {\bibinfo {volume} {32}},\ \bibinfo
  {pages} {109} (\bibinfo {year} {2021})}\BibitemShut {NoStop}%
\bibitem [{\citenamefont {Liu}\ \emph {et~al.}(2025{\natexlab{a}})\citenamefont
  {Liu}, \citenamefont {Wang}, \citenamefont {Zhang},\ and\ \citenamefont
  {Liu}}]{k300b7}%
  \BibitemOpen
  \bibfield  {author} {\bibinfo {author} {\bibfnamefont {G.~P.}\ \bibnamefont
  {Liu}}, \bibinfo {author} {\bibfnamefont {H.~L.}\ \bibnamefont {Wang}},
  \bibinfo {author} {\bibfnamefont {Z.~Z.}\ \bibnamefont {Zhang}},\ and\
  \bibinfo {author} {\bibfnamefont {M.~L.}\ \bibnamefont {Liu}},\ }\href
  {https://doi.org/10.1103/PhysRevC.111.024306} {\bibfield  {journal} {\bibinfo
   {journal} {Phys. Rev. C}\ }\textbf {\bibinfo {volume} {111}},\ \bibinfo
  {pages} {024306} (\bibinfo {year} {2025}{\natexlab{a}})}\BibitemShut
  {NoStop}%
\bibitem [{\citenamefont {Neufcourt}\ \emph {et~al.}(2019)\citenamefont
  {Neufcourt}, \citenamefont {Cao}, \citenamefont {Nazarewicz}, \citenamefont
  {Olsen},\ and\ \citenamefont {Viens}}]{3.c1}%
  \BibitemOpen
  \bibfield  {author} {\bibinfo {author} {\bibfnamefont {L.}~\bibnamefont
  {Neufcourt}}, \bibinfo {author} {\bibfnamefont {Y.}~\bibnamefont {Cao}},
  \bibinfo {author} {\bibfnamefont {W.}~\bibnamefont {Nazarewicz}}, \bibinfo
  {author} {\bibfnamefont {E.}~\bibnamefont {Olsen}},\ and\ \bibinfo {author}
  {\bibfnamefont {F.}~\bibnamefont {Viens}},\ }\href
  {https://doi.org/10.1103/PhysRevLett.122.062502} {\bibfield  {journal}
  {\bibinfo  {journal} {Phys. Rev. Lett.}\ }\textbf {\bibinfo {volume} {122}},\
  \bibinfo {pages} {062502} (\bibinfo {year} {2019})}\BibitemShut {NoStop}%
\bibitem [{\citenamefont {Neufcourt}\ \emph {et~al.}(2020)\citenamefont
  {Neufcourt}, \citenamefont {Cao}, \citenamefont {Giuliani}, \citenamefont
  {Nazarewicz}, \citenamefont {Olsen},\ and\ \citenamefont {Tarasov}}]{3.c2}%
  \BibitemOpen
  \bibfield  {author} {\bibinfo {author} {\bibfnamefont {L.}~\bibnamefont
  {Neufcourt}}, \bibinfo {author} {\bibfnamefont {Y.}~\bibnamefont {Cao}},
  \bibinfo {author} {\bibfnamefont {S.~A.}\ \bibnamefont {Giuliani}}, \bibinfo
  {author} {\bibfnamefont {W.}~\bibnamefont {Nazarewicz}}, \bibinfo {author}
  {\bibfnamefont {E.}~\bibnamefont {Olsen}},\ and\ \bibinfo {author}
  {\bibfnamefont {O.~B.}\ \bibnamefont {Tarasov}},\ }\href
  {https://doi.org/10.1103/PhysRevC.101.044307} {\bibfield  {journal} {\bibinfo
   {journal} {Phys. Rev. C}\ }\textbf {\bibinfo {volume} {101}},\ \bibinfo
  {pages} {044307} (\bibinfo {year} {2020})}\BibitemShut {NoStop}%
\bibitem [{\citenamefont {Liu}\ \emph {et~al.}(2021)\citenamefont {Liu},
  \citenamefont {Su}, \citenamefont {Liu}, \citenamefont {Danielewicz},
  \citenamefont {Xu},\ and\ \citenamefont {Ren}}]{k300b11}%
  \BibitemOpen
  \bibfield  {author} {\bibinfo {author} {\bibfnamefont {Y.~F.}\ \bibnamefont
  {Liu}}, \bibinfo {author} {\bibfnamefont {C.}~\bibnamefont {Su}}, \bibinfo
  {author} {\bibfnamefont {J.}~\bibnamefont {Liu}}, \bibinfo {author}
  {\bibfnamefont {P.}~\bibnamefont {Danielewicz}}, \bibinfo {author}
  {\bibfnamefont {C.}~\bibnamefont {Xu}},\ and\ \bibinfo {author}
  {\bibfnamefont {Z.~Z.}\ \bibnamefont {Ren}},\ }\href
  {https://doi.org/10.1103/PhysRevC.104.014315} {\bibfield  {journal} {\bibinfo
   {journal} {Phys. Rev. C}\ }\textbf {\bibinfo {volume} {104}},\ \bibinfo
  {pages} {014315} (\bibinfo {year} {2021})}\BibitemShut {NoStop}%
\bibitem [{\citenamefont {Wu}\ and\ \citenamefont {Zhao}(2020)}]{k300b14}%
  \BibitemOpen
  \bibfield  {author} {\bibinfo {author} {\bibfnamefont {X.~H.}\ \bibnamefont
  {Wu}}\ and\ \bibinfo {author} {\bibfnamefont {P.~W.}\ \bibnamefont {Zhao}},\
  }\href {https://doi.org/10.1103/PhysRevC.101.051301} {\bibfield  {journal}
  {\bibinfo  {journal} {Phys. Rev. C}\ }\textbf {\bibinfo {volume} {101}},\
  \bibinfo {pages} {051301} (\bibinfo {year} {2020})}\BibitemShut {NoStop}%
\bibitem [{\citenamefont {Tian}\ \emph {et~al.}(2025)\citenamefont {Tian},
  \citenamefont {Ma}, \citenamefont {Wu}, \citenamefont {Hu}, \citenamefont
  {Li},\ and\ \citenamefont {Wang}}]{9mgr-6mq7}%
  \BibitemOpen
  \bibfield  {author} {\bibinfo {author} {\bibfnamefont {J.~L.}\ \bibnamefont
  {Tian}}, \bibinfo {author} {\bibfnamefont {P.~F.}\ \bibnamefont {Ma}},
  \bibinfo {author} {\bibfnamefont {X.~H.}\ \bibnamefont {Wu}}, \bibinfo
  {author} {\bibfnamefont {M.~H.}\ \bibnamefont {Hu}}, \bibinfo {author}
  {\bibfnamefont {C.}~\bibnamefont {Li}},\ and\ \bibinfo {author}
  {\bibfnamefont {N.}~\bibnamefont {Wang}},\ }\href
  {https://doi.org/10.1103/9mgr-6mq7} {\bibfield  {journal} {\bibinfo
  {journal} {Phys. Rev. C}\ }\textbf {\bibinfo {volume} {112}},\ \bibinfo
  {pages} {064306} (\bibinfo {year} {2025})}\BibitemShut {NoStop}%
\bibitem [{\citenamefont {Niu}\ \emph {et~al.}(2019{\natexlab{b}})\citenamefont
  {Niu}, \citenamefont {Fang},\ and\ \citenamefont {Niu}}]{RBF1}%
  \BibitemOpen
  \bibfield  {author} {\bibinfo {author} {\bibfnamefont {Z.~M.}\ \bibnamefont
  {Niu}}, \bibinfo {author} {\bibfnamefont {J.~Y.}\ \bibnamefont {Fang}},\ and\
  \bibinfo {author} {\bibfnamefont {Y.~F.}\ \bibnamefont {Niu}},\ }\href
  {https://doi.org/10.1103/PhysRevC.100.054311} {\bibfield  {journal} {\bibinfo
   {journal} {Phys. Rev. C}\ }\textbf {\bibinfo {volume} {100}},\ \bibinfo
  {pages} {054311} (\bibinfo {year} {2019}{\natexlab{b}})}\BibitemShut
  {NoStop}%
\bibitem [{\citenamefont {Li}\ \emph {et~al.}(2025)\citenamefont {Li},
  \citenamefont {Wang}, \citenamefont {Li},\ and\ \citenamefont {Liu}}]{RBF2}%
  \BibitemOpen
  \bibfield  {author} {\bibinfo {author} {\bibfnamefont {T.}~\bibnamefont
  {Li}}, \bibinfo {author} {\bibfnamefont {N.}~\bibnamefont {Wang}}, \bibinfo
  {author} {\bibfnamefont {C.}~\bibnamefont {Li}},\ and\ \bibinfo {author}
  {\bibfnamefont {M.}~\bibnamefont {Liu}},\ }\href
  {https://doi.org/10.1103/h72z-3ytv} {\bibfield  {journal} {\bibinfo
  {journal} {Phys. Rev. C}\ }\textbf {\bibinfo {volume} {112}},\ \bibinfo
  {pages} {024306} (\bibinfo {year} {2025})}\BibitemShut {NoStop}%
\bibitem [{\citenamefont {Huang}\ \emph {et~al.}(2025)\citenamefont {Huang},
  \citenamefont {Chen}, \citenamefont {Jia}, \citenamefont {Liu}, \citenamefont
  {Ma},\ and\ \citenamefont {Zhang}}]{k300b6}%
  \BibitemOpen
  \bibfield  {author} {\bibinfo {author} {\bibfnamefont {Y.~M.}\ \bibnamefont
  {Huang}}, \bibinfo {author} {\bibfnamefont {J.~H.}\ \bibnamefont {Chen}},
  \bibinfo {author} {\bibfnamefont {J.~Y.}\ \bibnamefont {Jia}}, \bibinfo
  {author} {\bibfnamefont {L.~M.}\ \bibnamefont {Liu}}, \bibinfo {author}
  {\bibfnamefont {Y.~G.}\ \bibnamefont {Ma}},\ and\ \bibinfo {author}
  {\bibfnamefont {C.~J.}\ \bibnamefont {Zhang}},\ }\href
  {https://doi.org/10.1103/PhysRevC.111.034329} {\bibfield  {journal} {\bibinfo
   {journal} {Phys. Rev. C}\ }\textbf {\bibinfo {volume} {111}},\ \bibinfo
  {pages} {034329} (\bibinfo {year} {2025})}\BibitemShut {NoStop}%
\bibitem [{\citenamefont {Niu}\ and\ \citenamefont {Liang}(2022)}]{k300b15}%
  \BibitemOpen
  \bibfield  {author} {\bibinfo {author} {\bibfnamefont {Z.~M.}\ \bibnamefont
  {Niu}}\ and\ \bibinfo {author} {\bibfnamefont {H.~Z.}\ \bibnamefont
  {Liang}},\ }\href {https://doi.org/10.1103/PhysRevC.106.L021303} {\bibfield
  {journal} {\bibinfo  {journal} {Phys. Rev. C}\ }\textbf {\bibinfo {volume}
  {106}},\ \bibinfo {pages} {L021303} (\bibinfo {year} {2022})}\BibitemShut
  {NoStop}%
\bibitem [{\citenamefont {Bender}\ \emph {et~al.}(2003)\citenamefont {Bender},
  \citenamefont {Heenen},\ and\ \citenamefont {Reinhard}}]{4.e1}%
  \BibitemOpen
  \bibfield  {author} {\bibinfo {author} {\bibfnamefont {M.}~\bibnamefont
  {Bender}}, \bibinfo {author} {\bibfnamefont {P.~H.}\ \bibnamefont {Heenen}},\
  and\ \bibinfo {author} {\bibfnamefont {P.~G.}\ \bibnamefont {Reinhard}},\
  }\href {https://doi.org/10.1103/RevModPhys.75.121} {\bibfield  {journal}
  {\bibinfo  {journal} {Rev. Mod. Phys.}\ }\textbf {\bibinfo {volume} {75}},\
  \bibinfo {pages} {121} (\bibinfo {year} {2003})}\BibitemShut {NoStop}%
\bibitem [{\citenamefont {{Blaum}}(2006)}]{4.e2}%
  \BibitemOpen
  \bibfield  {author} {\bibinfo {author} {\bibfnamefont {K.}~\bibnamefont
  {{Blaum}}},\ }\href {https://doi.org/10.1016/j.physrep.2005.10.011}
  {\bibfield  {journal} {\bibinfo  {journal} {Phys. Rep.}\ }\textbf {\bibinfo
  {volume} {425}},\ \bibinfo {pages} {1} (\bibinfo {year} {2006})}\BibitemShut
  {NoStop}%
\bibitem [{\citenamefont {Wang}\ \emph {et~al.}(2014)\citenamefont {Wang},
  \citenamefont {Liu}, \citenamefont {Wu},\ and\ \citenamefont {Meng}}]{k6002}%
  \BibitemOpen
  \bibfield  {author} {\bibinfo {author} {\bibfnamefont {N.}~\bibnamefont
  {Wang}}, \bibinfo {author} {\bibfnamefont {M.}~\bibnamefont {Liu}}, \bibinfo
  {author} {\bibfnamefont {X.~Z.}\ \bibnamefont {Wu}},\ and\ \bibinfo {author}
  {\bibfnamefont {J.}~\bibnamefont {Meng}},\ }\href
  {https://doi.org/https://doi.org/10.1016/j.physletb.2014.05.049} {\bibfield
  {journal} {\bibinfo  {journal} {Phys. Lett. B}\ }\textbf {\bibinfo {volume}
  {734}},\ \bibinfo {pages} {215} (\bibinfo {year} {2014})}\BibitemShut
  {NoStop}%
\bibitem [{\citenamefont {Chabanat}\ \emph {et~al.}(1998)\citenamefont
  {Chabanat}, \citenamefont {Bonche}, \citenamefont {Haensel}, \citenamefont
  {Meyer},\ and\ \citenamefont {Schaeffer}}]{k6001}%
  \BibitemOpen
  \bibfield  {author} {\bibinfo {author} {\bibfnamefont {E.}~\bibnamefont
  {Chabanat}}, \bibinfo {author} {\bibfnamefont {P.}~\bibnamefont {Bonche}},
  \bibinfo {author} {\bibfnamefont {P.}~\bibnamefont {Haensel}}, \bibinfo
  {author} {\bibfnamefont {J.}~\bibnamefont {Meyer}},\ and\ \bibinfo {author}
  {\bibfnamefont {R.}~\bibnamefont {Schaeffer}},\ }\href
  {https://doi.org/https://doi.org/10.1016/S0375-9474(98)00180-8} {\bibfield
  {journal} {\bibinfo  {journal} {Nucl. Phys. A}\ }\textbf {\bibinfo {volume}
  {635}},\ \bibinfo {pages} {231} (\bibinfo {year} {1998})}\BibitemShut
  {NoStop}%
\bibitem [{\citenamefont {Ackley}\ \emph {et~al.}(1985)\citenamefont {Ackley},
  \citenamefont {Hinton},\ and\ \citenamefont {Sejnowski}}]{ACKLEY1985147}%
  \BibitemOpen
  \bibfield  {author} {\bibinfo {author} {\bibfnamefont {D.~H.}\ \bibnamefont
  {Ackley}}, \bibinfo {author} {\bibfnamefont {G.~E.}\ \bibnamefont {Hinton}},\
  and\ \bibinfo {author} {\bibfnamefont {T.~J.}\ \bibnamefont {Sejnowski}},\
  }\href {https://doi.org/https://doi.org/10.1016/S0364-0213(85)80012-4}
  {\bibfield  {journal} {\bibinfo  {journal} {Cogn. Sci.}\ }\textbf {\bibinfo
  {volume} {9}},\ \bibinfo {pages} {147} (\bibinfo {year} {1985})}\BibitemShut
  {NoStop}%
\bibitem [{\citenamefont {Geng}\ \emph
  {et~al.}(2025{\natexlab{b}})\citenamefont {Geng}, \citenamefont {Wang},
  \citenamefont {Chen}, \citenamefont {Gao}, \citenamefont {Frellsen},\ and\
  \citenamefont {Hauberg}}]{EBM}%
  \BibitemOpen
  \bibfield  {author} {\bibinfo {author} {\bibfnamefont {C.}~\bibnamefont
  {Geng}}, \bibinfo {author} {\bibfnamefont {J.}~\bibnamefont {Wang}}, \bibinfo
  {author} {\bibfnamefont {L.}~\bibnamefont {Chen}}, \bibinfo {author}
  {\bibfnamefont {Z.~Y.}\ \bibnamefont {Gao}}, \bibinfo {author} {\bibfnamefont
  {J.}~\bibnamefont {Frellsen}},\ and\ \bibinfo {author} {\bibfnamefont
  {S.}~\bibnamefont {Hauberg}},\ }\href
  {https://doi.org/10.1007/s11263-025-02460-0} {\bibfield  {journal} {\bibinfo
  {journal} {Int. J. Comput. Vis.}\ }\textbf {\bibinfo {volume} {133}},\
  \bibinfo {pages} {5898} (\bibinfo {year} {2025}{\natexlab{b}})}\BibitemShut
  {NoStop}%
\bibitem [{\citenamefont {Wang}\ \emph {et~al.}(2021)\citenamefont {Wang},
  \citenamefont {Huang}, \citenamefont {Kondev}, \citenamefont {Audi},\ and\
  \citenamefont {Naimi}}]{k1}%
  \BibitemOpen
  \bibfield  {author} {\bibinfo {author} {\bibfnamefont {M.}~\bibnamefont
  {Wang}}, \bibinfo {author} {\bibfnamefont {W.}~\bibnamefont {Huang}},
  \bibinfo {author} {\bibfnamefont {F.}~\bibnamefont {Kondev}}, \bibinfo
  {author} {\bibfnamefont {G.}~\bibnamefont {Audi}},\ and\ \bibinfo {author}
  {\bibfnamefont {S.}~\bibnamefont {Naimi}},\ }\href
  {https://doi.org/10.1088/1674-1137/abddaf} {\bibfield  {journal} {\bibinfo
  {journal} {Chin. Phys. C}\ }\textbf {\bibinfo {volume} {45}},\ \bibinfo
  {pages} {030003} (\bibinfo {year} {2021})}\BibitemShut {NoStop}%
\bibitem [{\citenamefont {Xie}\ \emph {et~al.}(2024)\citenamefont {Xie},
  \citenamefont {Wang}, \citenamefont {Wang}, \citenamefont {Gao},
  \citenamefont {Ju},\ and\ \citenamefont {Liu}}]{k300b12}%
  \BibitemOpen
  \bibfield  {author} {\bibinfo {author} {\bibfnamefont {J.~Z.}\ \bibnamefont
  {Xie}}, \bibinfo {author} {\bibfnamefont {K.~P.}\ \bibnamefont {Wang}},
  \bibinfo {author} {\bibfnamefont {C.}~\bibnamefont {Wang}}, \bibinfo {author}
  {\bibfnamefont {W.~Q.}\ \bibnamefont {Gao}}, \bibinfo {author} {\bibfnamefont
  {M.}~\bibnamefont {Ju}},\ and\ \bibinfo {author} {\bibfnamefont
  {J.}~\bibnamefont {Liu}},\ }\href
  {https://doi.org/10.1103/PhysRevC.109.064317} {\bibfield  {journal} {\bibinfo
   {journal} {Phys. Rev. C}\ }\textbf {\bibinfo {volume} {109}},\ \bibinfo
  {pages} {064317} (\bibinfo {year} {2024})}\BibitemShut {NoStop}%
\bibitem [{\citenamefont {Liu}\ \emph {et~al.}(2025{\natexlab{b}})\citenamefont
  {Liu}, \citenamefont {Tan}, \citenamefont {Wang}, \citenamefont {Gao},
  \citenamefont {Shang}, \citenamefont {Li},\ and\ \citenamefont {Xu}}]{CBPRC}%
  \BibitemOpen
  \bibfield  {author} {\bibinfo {author} {\bibfnamefont {J.}~\bibnamefont
  {Liu}}, \bibinfo {author} {\bibfnamefont {K.~Z.}\ \bibnamefont {Tan}},
  \bibinfo {author} {\bibfnamefont {L.}~\bibnamefont {Wang}}, \bibinfo {author}
  {\bibfnamefont {W.~Q.}\ \bibnamefont {Gao}}, \bibinfo {author} {\bibfnamefont
  {T.~S.}\ \bibnamefont {Shang}}, \bibinfo {author} {\bibfnamefont
  {J.}~\bibnamefont {Li}},\ and\ \bibinfo {author} {\bibfnamefont
  {C.}~\bibnamefont {Xu}},\ }\href {https://doi.org/10.1007/s41365-025-01792-3}
  {\bibfield  {journal} {\bibinfo  {journal} {Nucl. Sci. Tech.}\ }\textbf
  {\bibinfo {volume} {36}},\ \bibinfo {pages} {215} (\bibinfo {year}
  {2025}{\natexlab{b}})}\BibitemShut {NoStop}%
\bibitem [{\citenamefont {Huang}\ \emph {et~al.}(2017)\citenamefont {Huang},
  \citenamefont {Audi}, \citenamefont {Wang}, \citenamefont {Kondev},
  \citenamefont {Naimi},\ and\ \citenamefont {Xu}}]{k2}%
  \BibitemOpen
  \bibfield  {author} {\bibinfo {author} {\bibfnamefont {W.~J.}\ \bibnamefont
  {Huang}}, \bibinfo {author} {\bibfnamefont {G.}~\bibnamefont {Audi}},
  \bibinfo {author} {\bibfnamefont {M.}~\bibnamefont {Wang}}, \bibinfo {author}
  {\bibfnamefont {F.~G.}\ \bibnamefont {Kondev}}, \bibinfo {author}
  {\bibfnamefont {S.}~\bibnamefont {Naimi}},\ and\ \bibinfo {author}
  {\bibfnamefont {X.}~\bibnamefont {Xu}},\ }\href
  {https://doi.org/10.1088/1674-1137/41/3/030002} {\bibfield  {journal}
  {\bibinfo  {journal} {Chin. Phys. C}\ }\textbf {\bibinfo {volume} {41}},\
  \bibinfo {pages} {030002} (\bibinfo {year} {2017})}\BibitemShut {NoStop}%
\bibitem [{\citenamefont {Ni}\ and\ \citenamefont {Ren}(2012)}]{k3}%
  \BibitemOpen
  \bibfield  {author} {\bibinfo {author} {\bibfnamefont {D.~D.}\ \bibnamefont
  {Ni}}\ and\ \bibinfo {author} {\bibfnamefont {Z.~Z.}\ \bibnamefont {Ren}},\
  }\href {https://doi.org/https://doi.org/10.1016/j.nuclphysa.2012.08.006}
  {\bibfield  {journal} {\bibinfo  {journal} {Nucl. Phys. A}\ }\textbf
  {\bibinfo {volume} {893}},\ \bibinfo {pages} {13} (\bibinfo {year}
  {2012})}\BibitemShut {NoStop}%
\bibitem [{\citenamefont {Bao}\ \emph {et~al.}(2014)\citenamefont {Bao},
  \citenamefont {He}, \citenamefont {Zhao},\ and\ \citenamefont {Arima}}]{k4}%
  \BibitemOpen
  \bibfield  {author} {\bibinfo {author} {\bibfnamefont {M.}~\bibnamefont
  {Bao}}, \bibinfo {author} {\bibfnamefont {Z.}~\bibnamefont {He}}, \bibinfo
  {author} {\bibfnamefont {Y.~M.}\ \bibnamefont {Zhao}},\ and\ \bibinfo
  {author} {\bibfnamefont {A.}~\bibnamefont {Arima}},\ }\href
  {https://doi.org/10.1103/PhysRevC.90.024314} {\bibfield  {journal} {\bibinfo
  {journal} {Phys. Rev. C}\ }\textbf {\bibinfo {volume} {90}},\ \bibinfo
  {pages} {024314} (\bibinfo {year} {2014})}\BibitemShut {NoStop}%
\bibitem [{\citenamefont {Qi}\ \emph {et~al.}(2014)\citenamefont {Qi},
  \citenamefont {Andreyev}, \citenamefont {Huyse}, \citenamefont {Liotta},
  \citenamefont {{Van Duppen}},\ and\ \citenamefont {Wyss}}]{c1}%
  \BibitemOpen
  \bibfield  {author} {\bibinfo {author} {\bibfnamefont {C.}~\bibnamefont
  {Qi}}, \bibinfo {author} {\bibfnamefont {A.}~\bibnamefont {Andreyev}},
  \bibinfo {author} {\bibfnamefont {M.}~\bibnamefont {Huyse}}, \bibinfo
  {author} {\bibfnamefont {R.}~\bibnamefont {Liotta}}, \bibinfo {author}
  {\bibfnamefont {P.}~\bibnamefont {{Van Duppen}}},\ and\ \bibinfo {author}
  {\bibfnamefont {R.}~\bibnamefont {Wyss}},\ }\href
  {https://doi.org/https://doi.org/10.1016/j.physletb.2014.05.066} {\bibfield
  {journal} {\bibinfo  {journal} {Phys. Lett. B}\ }\textbf {\bibinfo {volume}
  {734}},\ \bibinfo {pages} {203} (\bibinfo {year} {2014})}\BibitemShut
  {NoStop}%
\bibitem [{\citenamefont {Olsen}\ and\ \citenamefont {Nazarewicz}(2019)}]{d1}%
  \BibitemOpen
  \bibfield  {author} {\bibinfo {author} {\bibfnamefont {E.}~\bibnamefont
  {Olsen}}\ and\ \bibinfo {author} {\bibfnamefont {W.}~\bibnamefont
  {Nazarewicz}},\ }\href {https://doi.org/10.1103/PhysRevC.99.014317}
  {\bibfield  {journal} {\bibinfo  {journal} {Phys. Rev. C}\ }\textbf {\bibinfo
  {volume} {99}},\ \bibinfo {pages} {014317} (\bibinfo {year}
  {2019})}\BibitemShut {NoStop}%
\bibitem [{\citenamefont {Tang}\ \emph {et~al.}(2025)\citenamefont {Tang},
  \citenamefont {Qian}, \citenamefont {Wan},\ and\ \citenamefont {Ren}}]{d2}%
  \BibitemOpen
  \bibfield  {author} {\bibinfo {author} {\bibfnamefont {S.~L.}\ \bibnamefont
  {Tang}}, \bibinfo {author} {\bibfnamefont {Y.~B.}\ \bibnamefont {Qian}},
  \bibinfo {author} {\bibfnamefont {T.}~\bibnamefont {Wan}},\ and\ \bibinfo
  {author} {\bibfnamefont {Z.~Z.}\ \bibnamefont {Ren}},\ }\href
  {https://doi.org/10.1103/l6ft-qvcs} {\bibfield  {journal} {\bibinfo
  {journal} {Phys. Rev. C}\ }\textbf {\bibinfo {volume} {112}},\ \bibinfo
  {pages} {014308} (\bibinfo {year} {2025})}\BibitemShut {NoStop}%
\bibitem [{\citenamefont {Hilton}\ \emph {et~al.}(2019)\citenamefont {Hilton},
  \citenamefont {Uusitalo}, \citenamefont {Sar\'en}, \citenamefont {Page},
  \citenamefont {Joss}, \citenamefont {AlAqeel}, \citenamefont {Badran},
  \citenamefont {Briscoe}, \citenamefont {Calverley}, \citenamefont {Cox},
  \citenamefont {Grahn}, \citenamefont {Gredley}, \citenamefont {Greenlees},
  \citenamefont {Harding}, \citenamefont {Herzan}, \citenamefont {Higgins},
  \citenamefont {Julin}, \citenamefont {Juutinen}, \citenamefont {Konki},
  \citenamefont {Labiche}, \citenamefont {Leino}, \citenamefont {Lewis},
  \citenamefont {Ojala}, \citenamefont {Pakarinen}, \citenamefont {Papadakis},
  \citenamefont {Partanen}, \citenamefont {Rahkila}, \citenamefont
  {Ruotsalainen}, \citenamefont {Sandzelius}, \citenamefont {Scholey},
  \citenamefont {Sorri}, \citenamefont {Sottili}, \citenamefont {Stolze},\ and\
  \citenamefont {Wearing}}]{d3}%
  \BibitemOpen
  \bibfield  {author} {\bibinfo {author} {\bibfnamefont {J.}~\bibnamefont
  {Hilton}}, \bibinfo {author} {\bibfnamefont {J.}~\bibnamefont {Uusitalo}},
  \bibinfo {author} {\bibfnamefont {J.}~\bibnamefont {Sar\'en}}, \bibinfo
  {author} {\bibfnamefont {R.~D.}\ \bibnamefont {Page}}, \bibinfo {author}
  {\bibfnamefont {D.~T.}\ \bibnamefont {Joss}}, \bibinfo {author}
  {\bibfnamefont {M.~A.~M.}\ \bibnamefont {AlAqeel}}, \bibinfo {author}
  {\bibfnamefont {H.}~\bibnamefont {Badran}}, \bibinfo {author} {\bibfnamefont
  {A.~D.}\ \bibnamefont {Briscoe}}, \bibinfo {author} {\bibfnamefont
  {T.}~\bibnamefont {Calverley}}, \bibinfo {author} {\bibfnamefont {D.~M.}\
  \bibnamefont {Cox}}, \emph {et~al.},\ }\href
  {https://doi.org/10.1103/PhysRevC.100.014305} {\bibfield  {journal} {\bibinfo
   {journal} {Phys. Rev. C}\ }\textbf {\bibinfo {volume} {100}},\ \bibinfo
  {pages} {014305} (\bibinfo {year} {2019})}\BibitemShut {NoStop}%
\bibitem [{\citenamefont {Ghiorso}\ \emph {et~al.}(1954)\citenamefont
  {Ghiorso}, \citenamefont {Thompson}, \citenamefont {Higgins}, \citenamefont
  {Harvey},\ and\ \citenamefont {Seaborg}}]{k8}%
  \BibitemOpen
  \bibfield  {author} {\bibinfo {author} {\bibfnamefont {A.}~\bibnamefont
  {Ghiorso}}, \bibinfo {author} {\bibfnamefont {S.~G.}\ \bibnamefont
  {Thompson}}, \bibinfo {author} {\bibfnamefont {G.~H.}\ \bibnamefont
  {Higgins}}, \bibinfo {author} {\bibfnamefont {B.~G.}\ \bibnamefont
  {Harvey}},\ and\ \bibinfo {author} {\bibfnamefont {G.~T.}\ \bibnamefont
  {Seaborg}},\ }\href {https://doi.org/10.1103/PhysRev.95.293} {\bibfield
  {journal} {\bibinfo  {journal} {Phys. Rev.}\ }\textbf {\bibinfo {volume}
  {95}},\ \bibinfo {pages} {293} (\bibinfo {year} {1954})}\BibitemShut
  {NoStop}%
\bibitem [{\citenamefont {Makii}\ \emph {et~al.}(2007)\citenamefont {Makii},
  \citenamefont {Ishii}, \citenamefont {Asai}, \citenamefont {Tsukada},
  \citenamefont {Toyoshima}, \citenamefont {Matsuda}, \citenamefont
  {Makishima}, \citenamefont {Kaneko}, \citenamefont {Toume}, \citenamefont
  {Ichikawa}, \citenamefont {Shigematsu}, \citenamefont {Kohno},\ and\
  \citenamefont {Ogawa}}]{k9}%
  \BibitemOpen
  \bibfield  {author} {\bibinfo {author} {\bibfnamefont {H.}~\bibnamefont
  {Makii}}, \bibinfo {author} {\bibfnamefont {T.}~\bibnamefont {Ishii}},
  \bibinfo {author} {\bibfnamefont {M.}~\bibnamefont {Asai}}, \bibinfo {author}
  {\bibfnamefont {K.}~\bibnamefont {Tsukada}}, \bibinfo {author} {\bibfnamefont
  {A.}~\bibnamefont {Toyoshima}}, \bibinfo {author} {\bibfnamefont
  {M.}~\bibnamefont {Matsuda}}, \bibinfo {author} {\bibfnamefont
  {A.}~\bibnamefont {Makishima}}, \bibinfo {author} {\bibfnamefont
  {J.}~\bibnamefont {Kaneko}}, \bibinfo {author} {\bibfnamefont
  {H.}~\bibnamefont {Toume}}, \bibinfo {author} {\bibfnamefont
  {S.}~\bibnamefont {Ichikawa}}, \emph {et~al.},\ }\href
  {https://doi.org/10.1103/PhysRevC.76.061301} {\bibfield  {journal} {\bibinfo
  {journal} {Phys. Rev. C}\ }\textbf {\bibinfo {volume} {76}},\ \bibinfo
  {pages} {061301} (\bibinfo {year} {2007})}\BibitemShut {NoStop}%
\bibitem [{\citenamefont {Chen}\ \emph {et~al.}(2025)\citenamefont {Chen},
  \citenamefont {Dong}, \citenamefont {Wang},\ and\ \citenamefont {Wu}}]{k10}%
  \BibitemOpen
  \bibfield  {author} {\bibinfo {author} {\bibfnamefont {B.~Y.}\ \bibnamefont
  {Chen}}, \bibinfo {author} {\bibfnamefont {J.~M.}\ \bibnamefont {Dong}},
  \bibinfo {author} {\bibfnamefont {Y.~Q.}\ \bibnamefont {Wang}},\ and\
  \bibinfo {author} {\bibfnamefont {G.~Q.}\ \bibnamefont {Wu}},\ }\href
  {https://doi.org/10.1088/1674-1137/ad8d4b} {\bibfield  {journal} {\bibinfo
  {journal} {Chin. Phys. C}\ }\textbf {\bibinfo {volume} {49}},\ \bibinfo
  {pages} {011001} (\bibinfo {year} {2025})}\BibitemShut {NoStop}%
\end{thebibliography}%
	
\end{document}